\documentclass[11pt,a4paper]{article}
\usepackage{amssymb} \usepackage{amsmath} \usepackage{graphicx}
\usepackage{epsfig,latexsym,color,xcolor}
\usepackage{ulem}
\usepackage{amstext}    
\usepackage{array} 
\usepackage{slashed}
\begin{document}

\def\Giulia{\bf\color{red}}
\def\bef{\begin{figure}}
\def\eef{\end{figure}}
\newcommand{\ans}{ansatz }
\newcommand{\be}[1]{\begin{equation}\label{#1}}
\newcommand{\beq}{\begin{equation}}
\newcommand{\ee}{\end{equation}}
\newcommand{\beqn}[1]{\begin{eqnarray}\label{#1}}
\newcommand{\eeqn}{\end{eqnarray}}
\newcommand{\bd}{\begin{displaymath}}
\newcommand{\ed}{\end{displaymath}}
\newcommand{\mat}[4]{\left(\begin{array}{cc}{#1}&{#2}\\{#3}&{#4}
\end{array}\right)}
\newcommand{\matr}[9]{\left(\begin{array}{ccc}{#1}&{#2}&{#3}\\
{#4}&{#5}&{#6}\\{#7}&{#8}&{#9}\end{array}\right)}

\newcommand{\matrr}[6]{\left(\begin{array}{cc}{#1}&{#2}\\
{#3}&{#4}\\{#5}&{#6}\end{array}\right)}
\newcommand{\cvb}[3]{#1^{#2}_{#3}}
\def\lsim{\raise0.3ex\hbox{$\;<$\kern-0.75em\raise-1.1ex
e\hbox{$\sim\;$}}}
\def\gsim{\raise0.3ex\hbox{$\;>$\kern-0.75em\raise-1.1ex
\hbox{$\sim\;$}}}
\def\abs#1{\left| #1\right|}
\def\simlt{\mathrel{\lower2.5pt\vbox{\lineskip=0pt\baselineskip=0pt
           \hbox{$<$}\hbox{$\sim$}}}}
\def\simgt{\mathrel{\lower2.5pt\vbox{\lineskip=0pt\baselineskip=0pt
           \hbox{$>$}\hbox{$\sim$}}}}
\def\unity{{\hbox{1\kern-.8mm l}}}
\newcommand{\eps}{\varepsilon}
\def\ep{\epsilon}
\def\ga{\gamma}
\def\Ga{\Gamma}
\def\om{\omega}
\def\omp{{\omega^\prime}}
\def\Om{\Omega}
\def\la{\lambda}
\def\La{\Lambda}
\def\al{\alpha}
\def\beq{\begin{equation}}
\def\eeq{\end{equation}}
\newcommand{\MyRed}{\color [rgb]{0.8,0,0}}
\newcommand{\MyGreen}{\color [rgb]{0,0.7,0}}
\newcommand{\MyBlue}{\color [rgb]{0,0,0.8}}
\def\GV#1{{\MyRed [GV: #1]}}
\def\MB#1{{\MyGreen [MB: #1]}}   
\def\AA#1{{\MyBlue [AA: #1]}}  
\newcommand{\sect}[1]{\setcounter{equation}{0}\section{#1}}
\renewcommand{\theequation}{\thesection.\arabic{equation}}
\newcommand{\ov}{\overline}
\renewcommand{\to}{\rightarrow}
\renewcommand{\vec}[1]{\mathbf{#1}}
\newcommand{\vect}[1]{\mbox{\boldmath$#1$}}
\def\tm{{\widetilde{m}}}
\def\mcirc{{\stackrel{o}{m}}}
\newcommand{\Dm}{\Delta m}
\newcommand{\dm}{\varepsilon}
\newcommand{\tanb}{\tan\beta}
\newcommand{\nbar}{\tilde{n}}
\newcommand\PM[1]{\begin{pmatrix}#1\end{pmatrix}}
\newcommand{\up}{\uparrow}
\newcommand{\down}{\downarrow}
\newcommand{\refs}[2]{eqs.~(\ref{#1})-(\ref{#2})}
\def\omE{\omega_{\rm Ter}}
\newcommand{\eqn}[1]{eq.~(\ref{#1})}
%

\newcommand{\DSUSY}{{SUSY \hspace{-9.4pt} \slash}\;}
\newcommand{\DCP}{{CP \hspace{-7.4pt} \slash}\;}
\newcommand{\mc}{\mathcal}
\newcommand{\gr}{\mathbf}
\renewcommand{\to}{\rightarrow}
\newcommand{\gtc}{\mathfrak}
\newcommand{\wh}{\widehat}
\newcommand{\br}{\langle}
\newcommand{\kt}{\rangle}


\def\lsim{\mathrel{\mathop  {\hbox{\lower0.5ex\hbox{$\sim$}
\kern-0.8em\lower-0.7ex\hbox{$<$}}}}}
\def\gsim{\mathrel{\mathop  {\hbox{\lower0.5ex\hbox{$\sim$}
\kern-0.8em\lower-0.7ex\hbox{$>$}}}}}

\def\nn{\\  \nonumber}
\def\de{\partial}
\def\brf{{\mathbf f}}
\def\bbf{\bar{\bf f}}
\def\bF{{\bf F}}
\def\bbF{\bar{\bf F}}
\def\bA{{\mathbf A}}
\def\bB{{\mathbf B}}
\def\bG{{\mathbf G}}
\def\bI{{\mathbf I}}
\def\bM{{\mathbf M}}
\def\bY{{\mathbf Y}}
\def\bX{{\mathbf X}}
\def\bS{{\mathbf S}}
\def\bb{{\mathbf b}}
\def\bh{{\mathbf h}}
\def\bg{{\mathbf g}}
\def\bla{{\mathbf \la}}
\def\bmu{\mathbf m }
\def\by{{\mathbf y}}
\def\bmu{\mbox{\boldmath $\mu$} }
\def\bsig{\mbox{\boldmath $\sigma$} }
\def\bunity{{\mathbf 1}}
\def\cA{{\cal A}}
\def\cB{{\cal B}}
\def\cC{{\cal C}}
\def\cD{{\cal D}}
\def\cF{{\cal F}}
\def\cG{{\cal G}}
\def\cH{{\cal H}}
\def\cI{{\cal I}}
\def\cL{{\cal L}}
\def\cN{{\cal N}}
\def\cM{{\cal M}}
\def\cO{{\cal O}}
\def\cR{{\cal R}}
\def\cS{{\cal S}}
\def\cT{{\cal T}}
\def\eV{{\rm eV}}
%

\large
 \begin{center}
 {\Large \bf  Soft gravitational radiation from ultra-relativistic collisions \\ at sub- and sub-sub-leading order}

 \end{center}

 \vspace{0.1cm}

 \vspace{0.1cm}
 \begin{center}
{\large Andrea Addazi}\footnote{E-mail: \,  andrea.addazi@qq.com} \\

{\it Center for Field Theory and Particle Physics \& Department of Physics, \\ Fudan University, 200433 Shanghai, China}
\end{center}

  \begin{center}
{\large Massimo Bianchi}\footnote{E-mail: \, massimo.bianchi@roma2.infn.it}
\\
{\it Dipartimento di Fisica, Universit\`a di Roma Tor Vergata, \\
I.N.F.N. Sezione di Roma Tor Vergata, \\
Via della Ricerca Scientifica, 1 00133 Roma, Italy}
\end{center}

  \begin{center}
{\large Gabriele Veneziano}\footnote{E-mail: \, Gabriele.Veneziano@cern.ch}
\\
{\it Theory Department, 
CERN, CH-1211 Geneva 23, Switzerland  \\ Coll\'ege de France, 11 place M. Berthelot, 75005 Paris, France}
\end{center}

\vspace{1cm}
\begin{abstract}
\small
Using  soft-graviton theorems  a well-known zero-frequency limit (ZFL) for the gravitational radiation  flux $dE^{GW}/d \omega$  is re-derived and extended to order ${{\cal O}}(\omega)$ and ${{\cal O}}(\omega^2)$ for  arbitrary massless multi-particle collisions. The (angle-integrated, unpolarized) ${{\cal O}}(\omega)$ correction to the flux turns out to be absent in the case of two-particle elastic collisions. The ${{\cal O}}(\omega^2)$ correction
is instead non-vanishing and takes a simple general expression which is then applied to bremsstrahlung from two-particle elastic collisions. For a tree-level process the outcome is finite and consistent with expectations. Instead,  if the tree-level form of the soft theorems is used at sub-sub-leading order even when the elastic amplitude needs an all-loop (eikonal) resummation, an unphysical infrared singularity occurs. Its origin can be traced to the infinite Coulomb phase of gravitational scattering in four dimensions.  We briefly discuss how to get rid, in principle, of the unwanted divergences and indicate --without carrying out-- a possible procedure to find the proper correction to the naive soft theorems.
Nevertheless, if a simple recipe recently proposed for handling these divergences is adopted, we find surprisingly good agreement with results obtained independently via the eikonal approach to transplanckian-energy scattering at large (small) impact parameter (deflection angle), where such Coulomb divergences explicitly cancel out.

\end{abstract}
\vspace{1cm}
\noindent{\bfseries\large\sffamily{Preprints:}}  CERN-TH-2018-269
\newpage

\section{Introduction}

A well known result in gravitational bremsstrahlung is a general formula  \cite{Smarr:1977fy} for the (finite) zero-frequency  limit (ZFL) of the spectrum of gravitational waves $dE^{GW}{/}{d\omega}$ emitted in a generic process as $\omega \rightarrow 0$. Such a formula can be derived either by classical considerations or by using the leading-order soft-graviton theorems going back to the classic paper of Weinberg's \cite{Weinberg} and others before \cite{others3, others1} and after \cite{others2, others4, others5, KF} him.

Interest in soft theorems has been revived in recent years with much work \cite{Donoghue:1999qh, 9, 6, 10, 11,Bianchi:2014gla, Bern:2014vva,DiVecchia:2015oba, DiVecchia:2015jaq, Bianchi:2016viy,Bianchi:2015lnw, Bianchi:2016tju,Guerrieri:2017ujb, Sen:2017xjn, Sen:2017nim, Laddha:2017ygw, Chakrabarti:2017zmh} related the sub and sub-sub-leading corrections to the leading soft term, to their (lack of) universality and to their connection \cite{5,7,8,Strominger,SHP, zhiboedov} to the Bondi-Metzner-Van der Burg-Sachs (BMS)\cite{BMS} group of isometries of asymptotically-flat space-times.

At the same time, new techniques have been developed to compute graviton amplitudes through their connection to gauge theory amplitudes \cite{Dunbar:1994bn, Bern:1998sv}.
While the main thrust in those papers has been towards understanding the ultraviolet structure of quantum gravity or supergravity some attention has also been devoted to the infrared (IR) divergences and their cancellation in sufficiently inclusive quantities.

Although at the time of Weinberg's paper these problems looked  highly academic, the recent direct observations of GW from BH and NS mergers \cite{GWobs} can make the issue of gravitational radiation in highly energetic collisions relevant even outside the merger regimes (see e.g Ref. \cite{Damour18}). 

An obvious question to ask is whether the new developments in soft theorems can be used to extend the predictions about $dE^{GW}{/}d\omega$ away from the ZFL.
This program has been recently undertaken by Laddha and Sen \cite{Laddha:2018rle} with very encouraging results. Because of the IR problems with gravity in $D=4$ most of the work in \cite{Laddha:2018rle} has focussed on $D > 4$ but, very recently an extension to the physically relevant case has been attempted by Sahoo and Sen \cite{Sahoo:2018lxl}.

In an unrelated  development the problem of high-energy gravitational scattering has been studied, particularly since the late eighties. In that case the original motivations were somewhat different: studying gravitational scattering at transplanckian energies \cite{Amati:1987wq, Muzinich:1987in, tHooft:1987vrq, Amati:1987uf} (see also \cite{Gross:1987kza, Gross:1987ar, Mende:1989wt}) can shed light on the information puzzle in regimes in which, at least classically, a black hole should form \cite{Eardley:2002re, Kohlprath:2002yh, Yoshino:2004mm, Giddings:2004xy} and then seen evaporate via the Hawking process \cite{Hawking:1974sw}. Another motivation, in the context of string theory, was to study new effects  due to the finite size of strings  such as tidal excitations \cite{Amati:1987wq, Amati:1987uf, Giddings:2006vu} or possible short-distance modifications of gravity \cite{Amati:1988tn, D'Appollonio:2015gpa}.  

That program is still ongoing: in spite some definite progress (see e.g. \cite{Amati:1990xe, Amati:2007ak, Veneziano:2004er} and references therein) constructing a unitary $S$-matrix in the black hole formation regime, whereby the impact parameter $b$ is smaller than the Schwarzschild radius $R_S=2G\sqrt{s}$ at the given CM energy $E_{CM}=\sqrt{s}$, has resisted so far all attempts. For this reason the attention shifted, at least momentarily, to the study of gravitational bremsstrahlung, an unavoidable phenomenon that should already occur well before the collapse regime is attained. Indeed, in the ultra-relativistic regime, gravitational radiation occurs already at next to leading order in the expansion of the eikonal phase $\delta(s,b)$ in Einstein's deflection angle $\theta_E = 2 R_S/b$. This is why the bremsstrahlung problem was first addressed at leading ({\it i.e.} $\theta_E^2$) order.

Two very distinct methods have been followed: a classical one \cite{Gruzinov:2014moa}, using Huygens principle in the Fraunhofer approximation and a second one \cite{Ciafaloni:2015xsr} (see also \cite{Ciafaloni:2015vsa}, \cite{Ciafaloni:2016nul}) following the semiclassical approximations implicit in the Amati-Ciafaloni-Veneziano (ACV) approach. Amazingly, the two approaches gave the same result for $dE^{GW}{/}d\omega$ as long as $\hbar \omega \ll \sqrt{s}$.
The features of the spectrum are quite interesting: they are consistent with the ZFL for $\omega<1/b$, deviate only  logarithmically from the ZFL constant for $1/b<\omega<1/R_S$, and finally drop above the ``Hawking frequency" $1/R_S$. This intriguing inverse proportionality between the initial transplanckian energy $\sqrt{s}$ and the characteristic energy of the final gravitons $\hbar/R_s$ is also the leitmotif of other studies \cite{Dvali:2014ila, Addazi:2016ksu} \footnote{There is still an issue about the high-frequency limit of the spectrum not being sufficiently suppressed. This appears to be related to some of the approximations breaking down at high frequency: they are of no concern for the present low-frequency investigation.}.

The basic motivation of this work is to build a bridge between the soft theorem (valid at sufficiently small $\omega$ but for arbitrary kinematics) and the results of \cite{Gruzinov:2014moa, Ciafaloni:2015xsr,Ciafaloni:2015vsa, CCV18} (valid at small-deflection angle but presumably in a wider frequency range)  and check their mutual consistency in the overlap of the two regimes where both of them can be  trusted.
   
To this end we will extend the calculation of the ZFL  to the next two sub-leading terms by taking advantage of the known sub-leading corrections to the soft graviton theorems.
Since in this paper we consider directly the physical case of $4$-dimensional spacetime, we will have to face some infrared problems that would be absent in higher dimensions. 
Let us start therefore by outlining  the assumptions of our approach and their possible limitations.

Consider a generic process $i \rightarrow f$ in which we have excluded both real and virtual soft-graviton processes/corrections\footnote{For the sake of simplicity we consider the case of Standard Model singlet particles so that we do not have to worry about soft gauge bosons.}. Let us denote by ${\mathcal S}^{(0)}_{fi}$ the bare (uncorrected) $S$-matrix for the process:
\be{S0}
{\mathcal S}^{(0)}_{fi} = \langle f | {\mathcal S}^{(0)} | i \rangle\, . 
\ee

We will assume that, to the desired order in the low-energy expansion, the inclusion of virtual and real soft-gravitons dresses the bare $S$-matrix with a coherent-state operator
\beqn{S}
&& {\mathcal S}^{(0)} \rightarrow {\mathcal S} = \exp \left( \int \frac{d^3 q}{\sqrt{2 \omega}} ({\lambda}^*_q a^{\dagger}_q - {\lambda}_q a_q)  \right) {\mathcal S}^{(0)}\\
&\equiv& \exp \left( - \frac12 \int_{\lambda}^{\Lambda} \frac{d^3 q}{2 \omega} |{\lambda}_q|^2  \right)
 \exp \left( \int_{\lambda}^{\Lambda} \frac{d^3 q}{\sqrt{2 \omega}} {\lambda}^*_q a^{\dagger}_q\right)  \exp \left(- \int_{\lambda}^{\Lambda} \frac{d^3 q}{\sqrt{2 \omega}}  {\lambda}_q a_q \right) {\mathcal S}^{(0)}\, .  \nonumber 
\eeqn
In the last term of (\ref{S}), which specifies the definition of the coherent state, $a_q, a^{\dagger}_q$ are destruction and creation operators for a (soft) graviton of momentum $q$ and maximal energy $\Lambda$. $\Lambda \ll E$ (with $E$ the characteristic energy-scale of the process) defines what one means by ``soft", $\lambda$ is an infrared cutoff, and ${\lambda}_q$ is a process-dependent ``function" of $q$ to be discussed below. 

If the initial state contains no soft gravitons the cross section for emitting 
any number of soft-gravitons (sgr) with a maximal total energy $\Delta E$ (basically identified with the energy resolution of the detector)  will exhibit the well-known cancellation of virtual and real soft divergences in the form:
\be{total}
\sum_{{\rm sgr(\Delta E)}} | \langle f; {\rm sgr}|  {\mathcal S} |i \rangle|^2 \sim  |\langle f |  {\mathcal S}^{(0)} | i \rangle|^2 \exp \left( -  \int_{\Delta E}^{E} \frac{d^3 q}{2 \omega} |{\lambda}_q|^2  \right)
\ee 
with a characteristic dependence on $\Delta E$.  We have assumed that $\Delta E$  is negligible w.r.t. the total energy $E$ of the process (otherwise there are correction terms, see \cite{Weinberg}).

Let us now consider the expectation value of the energy carried by the soft-gravitons in the process at hand. It will be given by the expectation value (in the appropriate coherent state) of the gravitational-energy operator $ \int d^3 q \hbar \omega a^{\dagger}_q a_q$: 
\beqn{exp} 
&& \langle 0|  \langle i |  {\mathcal S}^{\dagger} |f \rangle \int  d^3 q \hbar \omega a^{\dagger}_q a_q \langle f|  {\mathcal S} | i \rangle  |0 \rangle = \exp \left( -  \int_{\lambda}^{\Lambda}  \frac{d^3 q}{2 \omega} |{\lambda}_q|^2  \right)   \nonumber \\
&\times&   \langle 0|  \exp \left( \int_{\lambda}^{\Lambda} \frac{d^3 q}{\sqrt{2 \omega}} {\lambda}_q a_q\right)   \int d^3 q \hbar  \omega a^{\dagger}_q a_q  \exp \left( \int_{\lambda}^{\Lambda} \frac{d^3 q}{\sqrt{2 \omega}} {\lambda}^*_q a^{\dagger}_q\right) |0 \rangle |\langle f |  {\mathcal S}^{(0)} | i \rangle|^2 \nonumber \\
&& = \int_{\lambda}^{\Lambda} \frac{d^3 q}{2 \omega} \hbar  \omega |{\lambda}_q|^2 \sum_{{\rm sgr(\Lambda)}} | \langle f; {\rm sgr}|  {\mathcal S} |i \rangle|^2
\eeqn
where $\omega \equiv |q|$, the vacuum states refer to the soft-graviton Fock space, and we have used a well-known property of coherent states.

Equation (\ref{exp}) is the basic equation to be exploited in order to find the connection between soft-graviton theorems and the spectrum of gravitational radiation from a given process. Indeed, factoring out the probability for $i \rightarrow f + {\rm soft~gr.}$,  it follows from (\ref{exp}) that:
\be{spectrum}
 \frac{dE^{GW}(i \rightarrow f)}{d^3 q} = \frac{\hbar}{2}  |{\lambda}_q^{(i \rightarrow f)}|^2
\ee
where we have emphasized that $\lambda_q$ depends on the process $i \rightarrow f$ under consideration.\footnote{Our procedure should be justified if $\Delta E \ll E$ but much larger than the typical energy $\hbar \omega$ of the gravitons whose spectrum we wish to compute. A more rigorous framework should be provided by the Kulish-Faddeev prescription \cite{KF}. Our assumption basically amounts to assuming that such a prescription extends to gravity up to the sub-sub-leading term.}

Expanding ${\lambda}_q$ in power of $q$ will give the corresponding expansion of the GW spectrum.
The leading term in $ |{\lambda}_q^{(i \rightarrow f)}|^2$ goes like $\omega^{-2}$ which, given the phase space volume $d^3 q$, leads to a constant $dE^{GW}{/}{d\omega}$, the above mentioned ZFL. Naively, the two next-to-leading corrections will correspond to terms in $dE^{GW}{/}{d\omega}$ going to zero as $\omega$ and $\omega^2$  up to possible logarithmic enhancements to be discussed later.

There is, however, a subtlety concerning the sub-leading contributions to ${\lambda}_q$ which can already be seen at tree level.
Unlike the leading term the non-leading ones contain differential operators acting on the bare amplitude. 
When we eventually take the square modulus of the graviton amplitude we have to be careful about which  bare amplitude ($ {\mathcal S}$ or $ {\mathcal S}^{\dagger}$)  these operators act on.

A more important issue is what happens to these differential operators when we replace the bare amplitude with the one dressed by the insertion of soft gravitons.
It is well known that, in $D=4$, the naive operators should suffer modifications because of IR divergences \cite{Bern:2014oka}. While the final answer should be free of such divergences one expects finite correction for the sub and sub-sub-leading terms.

There are actually two kinds of IR divergences that can induce finite modifications to the naive recipe.
The first is directly related to the cancellation of real and virtual divergences we have already mentioned. It has been discussed, in particular, in \cite{Bern:2014oka}.
Once the cancellation is achieved some new finite contributions may be left. To our knowledge, no systematic analysis of these finite corrections has been given so far\footnote{One of us (GV) would like to thank Z. Bern for an interesting discussion about this point.}. In Sect. 6 we will argue that, under the already mentioned condition $\Delta E \gg \hbar \omega$, these corrections should be unimportant.

The second kind of divergences is related to the infinite Coulomb phase due to the long-range exchange of massless gravitons. This can also be seen as a divergent time delay in $D=4$ which might be cured by properly defining the asymptotic states as in \cite{KF}. This case has been recently addressed in a series of interesting papers by Laddha and Sen \cite{Laddha:2017ygw, Laddha:2018rle} and by Sahoo and Sen  \cite{Sahoo:2018lxl} and some recipe about how to deal with them has been proposed. In this case the infinity should be unobservable and simply disappear from any physical observable (such as the energy flux). We shall see, however, that the naive soft theorems at sub-leading level do not eliminate the divergence.
An appropriate modification of the soft theorems will remove it but, as pointed out in \cite{ Laddha:2018rle} may also leave behind some finite (logarithmic) corrections.
Their existence was recently confirmed in \cite{CCV18} within the eikonal approach.

For the time being we will proceed as if the naive prescription suffers no modification even at $D=4$. The need and possible form of the modifications will be discussed in Section 6.

The emission of  soft gravitons at tree level  is governed, at the first three leading orders, by the universal behavior \cite{5, Bianchi:2014gla, Bern:2014vva}  

  \be{softoper}
 {\cal M}_{N+1}(p_i; q) \approx \kappa \sum_{i=1}^N \left[ {p_i h p_i \over  qp_i}  + {p_i h J_i q \over qp_i} -
  {q J_i h J_i q \over 2 qp_i}\right] {\cal M}_{N}(p_i) \equiv S_0 + S_1 + S_2 \, , 
 \ee
 where $\kappa^2 = 8 \pi G$, $q^\mu$ and $h_{\mu\nu}$ denote the momentum and the polarization of the soft graviton, while $J^{\mu\nu}_i = p^\mu_i \partial/\partial p^i_\nu - p^\nu_i \partial/\partial p^i_\mu$ denotes the angular momentum operator\footnote{To compare with the literature note that we have omitted a factor $i$ in the definition of $J$. All our momenta, including the soft graviton's, are taken to be incoming. We also use the $-+++$ Minkowski  metric. Finally, one should not confuse the soft operators $S_i$ in (\ref{softoper}) with the $S$-matrix introduced in Eqs.(1.1)-(1.2) and denoted by ${\mathcal S}$.}  of the $i$th `hard' particle. In this paper, for simplicity, we  consider the collision of spin-less particles for which an additional spin term in $J^{\mu\nu}_i $ is absent. We will also take the massless limit, which is theoretically interesting, should be free of collinear divergences and, as a result,  should apply  to the case of ultra-relativistic collisions.
 
 An important property of (\ref{softoper}) is its gauge invariance {\it i.e.} the fact that it is invariant under the replacement:
 
 \be{gi}
 h_{\mu\nu} \rightarrow h_{\mu\nu} + q_\mu \xi_\nu + q_\nu \xi_\mu
 \ee
 which can be easily checked to hold thanks to conservation of linear and angular momentum.
 
 It is straightforward to connect the ``function'' $\lambda_q $ of (\ref{S}) to the soft operators $S_i$ of (\ref{softoper}). This
allows to compute the graviton spectrum (and thus the GW energy spectrum) to leading, next to leading and next to next to leading accuracy.
Although we shall start with the fully differential distribution, in this paper we will consider the frequency spectrum after integration over the direction of the emitted gravitational wave.
We will first review the leading order computation,  and then consider the  sub-leading corrections. In the rest of the paper we shall assume that Eq. (\ref{softoper}) continues to be valid beyond tree level. Such an assumption (for the non-leading term) will have to be checked case by case.

\vspace{0.7cm}
The plan of the paper is as follows. 
In Section 2, we will rederive the zero-frequency limit of the GW spectrum and point out the possible presence, in Weinberg's soft theorem, of sub-leading terms. 
In Section 3, we present the general computation of the ${{\cal O}}(\omega)$ correction to the GW spectrum. Sub-section 3.1 deals with the general case while 
in Sub-sections 3.2 and 3.3  we will show, by two different methods, that the ${{\cal O}}(\omega)$ correction to the two particle elastic collision is absent.
In Section 4, we present the ${{\cal O}}(\omega^{2})$ corrections to the GW spectrum in the general case. In Section 5.1 we specialize again to the case of 2-body scattering
and then we apply it to two particular cases:  in Subsection 5.2 to a tree-level gravitational scattering and,  in Subsections 5.3, to  the leading eikonal resummation of trans-planckian scattering at small deflection angle pointing out a possible infrared problem for the latter case.
In Section 6 we discuss the issue of IR divergences, their elimination, and the possible finite terms that originate from them.
In Section 7, we conclude and outline some directions for future investigation. 
Some details of the computations are relegated to Appendices A and B. Appendix C contains an {\it ab-initio} calculation of graviton emission from the tree-level gravitational scattering of scalar particles in ${\cal N}=8$ supergravity, analogous to the process discussed in Subsection  5.2, showing complete agreement between the two calculations.


\section{The GW spectrum at leading order}
\setcounter{equation}{0}

To leading order, the emission of real soft gravitons (gravi-strahlung) typically contributes a factor
$({\Delta E}/{\lambda})^{B_{0}}$ to the inclusive cross section, where $\Delta E$ is the maximal energy allowed in soft gravitons and $\lambda$ is an infrared cutoff.
  
The dominant behaviour, which is universal, {\it i.e.} valid at any order in perturbation theory in any consistent quantum theory of gravity such as String Theory, reads \cite{Weinberg}  \be{softadd}
\left\vert{\cal M}_{N+1}(p_i; q)\right\vert^2 = 8 \pi G \sum_{s=\pm 2} \left\vert\sum_{i=1}^N {p_i h_s p_i \over  qp_i}\right\vert^2 \left\vert{\cal M}_{N}(p_i)\right\vert^2 \, .
\ee
The sum over polarisations/helicities $s=\pm 2$, with $h^{-s}_{\mu\nu} = (h^{s}_{\mu\nu})^* $, produces a transverse traceless bi-symmetric tensor
\be{tttensor}
\sum_{s=\pm 2} h^s_{\mu\nu} h^{-s}_{\rho\sigma}  = \Pi_{\mu\nu,\rho\sigma} = {1\over 2} 
\left( {\pi}_{\mu\rho}{\pi}_{\nu\sigma} + {\pi}_{\mu\sigma}{\pi}_{\nu\rho} - {\pi}_{\mu\nu}{\pi}_{\rho\sigma}\right)\, ,
\ee 
where ${\pi}_{\mu\nu}= \eta_{\mu\nu} - q_\mu \bar{q}_\nu -q_\nu \bar{q}_\mu$ with $\bar{q}^2=0$ and $\bar{q} q = 1$. Luckily most of the terms are irrelevant thanks to momentum conservation, which, at leading order in $q$, reads $\sum_i  p_i = 0$, and to the mass-shell condition $p_i^2=0$. A straightforward calculation then gives
\be{sumij}\sum_{i,j} {p^\mu_i p^\nu_i \over qp_i} \Pi_{\mu\nu,\rho\sigma} {p^\rho_j p^\sigma_j \over qp_j} = 
\sum_{i,j}  {(p_i p_j)^2  \over qp_i qp_j } \, . \ee

Integration over the soft (final) light-like momentum $-q= (|q|, -\vect{q}) = |q|(1, -\vect{n})$ produces the well-known infrared divergent result \cite{Weinberg, Addazi:2016ksu}:
\be{Bagain}
B_0 = \frac{8\pi G}{\hbar} \int {d^3q\over 2|q| (2\pi)^3} \sum_{i,j}  {(p_i p_j)^2  \over qp_i qp_j } = - {2G\over \pi \hbar} \log{\Lambda\over \lambda}  \sum_{i,j}  (p_i p_j) \log {|p_i p_j|\over \mu^2}\, ,
\ee
where $\Lambda$ is a characteristic energy-scale of the process. Note that  the mass scale $\mu$ can be chosen at will since $\log\mu$ drops out thanks to momentum conservation to leading order in $q$.

The infrared divergence appearing in (\ref{Bagain}) corresponds to a bremsstrahlung spectrum for the number density ${dN_0}{/}{d\omega}$ given by
\be{B0Nspectrum}
\frac{d B_0}{d \omega} = \frac{8\pi G}{\hbar} \int {d^3q~ \delta( |q|-\omega) \over 2|q| (2\pi)^3}\sum_{i,j}  {(p_i p_j)^2  \over qp_i qp_j } = - {2G\over \pi \hbar \omega}  \sum_{i,j}  (p_i p_j) \log {|p_i p_j|\over \mu^2} \, , 
\ee
and thus to an energy spectrum:
\be{B0Espectrum}
\frac{d E_0^{GW}}{d \omega} =  \hbar \omega \frac{d N_0}{d \omega} = - {2G\over \pi }   \sum_{i,j}  (p_i p_j) \log {|p_i p_j|\over \mu^2} \, ,
\ee
which is nothing but the constant ZFL of the gravitational radiation associated with the specific process under consideration. 

 Noting that in the sum over $i \ne j$ each pair $i,j$ is counted twice, we can replace $- (p_i p_j)$ by $-(p_i + p_j)^2$ i.e. by the Mandelstam variable associated with each distinct particle pair. In the special case of a $2\rightarrow 2$ process an extra factor of two follows from the fact that two distinct pairs share the same Mandelstam variable ($s,t$, or $u$). The final formula in that case reads:
\be{ZFL4p}
\frac{d E^{GW}}{d \omega}(\omega = 0) = \frac{4 G}{\pi} \left( s \log s + t \log(-t) + u \log(-u) \right)\; .
\ee
In the small-$t$ (deflection angle $\theta_s$) limit this gives:
\be{ZFL4psa}
\frac{d E^{GW}}{d \omega}(\omega = 0) = \frac{Gs}  {\pi} \theta_s^2 \log (4 e \, \theta_s^{-2})\; , ~~  \theta_s = 4 G \sqrt{s}/b \; ,
\ee
in agreement with recent classical and quantum calculations \cite{Gruzinov:2014moa, Ciafaloni:2015vsa, Ciafaloni:2015xsr}.

In the next two sections we will extend the result (\ref{B0Espectrum}) to the first two next to leading orders in $\omega$. 
However, in order to have the complete result to sub-leading order one has to take into account possible sub-leading contributions already contained in the above calculation.
Indeed, to subleading order the momenta $p_i$ appearing on the l.h.s. of (\ref{softadd}) differ from those appearing on the r.h.s. In (\ref{Bagain}), (\ref{B0Nspectrum}), (\ref{B0Espectrum}) we have also neglected higher order terms:  in particular, if momentum conservation were written as $\sum_ip_i=-q$, one would get, instead of (\ref{sumij}):
\be{sumij1}\sum_{i,j} {p^\mu_i p^\nu_i \over qp_i} \Pi_{\mu\nu,\rho\sigma} {p^\rho_j p^\sigma_j \over qp_j} = 
\sum_{i,j}  {(p_i p_j)^2  \over qp_i qp_j } - 4  \, . 
\ee
An even more subtle point is that sub-leading corrections to eq. (\ref{B0Nspectrum}) depend on exactly which variables are kept fixed while one integrates over angles.
Physically, we certainly want to keep the center-of-mass energy $\sqrt{s_{12}}$ fixed and, by fixing $\omega$, also $s_{34}$ is kept constant. However, there is some ambiguity on the 3rd variable one wishes to keep fixed\footnote{Recall that the 5-point function depends of five independent variables, hence, after angular integration, the energy spectrum should depend on 3 variables.}.

We will come back to such terms after discussing the other non-leading corrections stemming from (\ref{softoper}).


\section{The ${{\cal O}}(\omega)$ correction to the GW spectrum }
\setcounter{equation}{0}

The first sub-leading correction arises from the interference between $S_0$ and $S_1$ of eq. (\ref{softoper})
\be{B1}
B_1 = 8\pi G \int {d^3q\over 2|q| (2\pi)^3} \sum_{i,j}  \sum_{s=\pm 2} \left[{(p_i h^s p_i)(p_j h^{(-s)}J_j q)  \over qp_i qp_j } + (i\leftrightarrow j) \right]\, .
\ee

After some tedious but straightforward algebra, the sum over helicities of the emitted graviton produces
\be{Sigma}
\Sigma = \sum_{i,j} p_i^\mu p_i^\nu p_j^\rho (J_jq)^\sigma \Pi_{\mu\nu,\rho\sigma} + (i\leftrightarrow j) = \sum_{i,j} 
{p_ip_j \over qp_i qp_j } [p_iJ_jq + p_jJ_iq] \, ,
\ee
where Poincar\'e invariance has been taken into account to set $\sum_i p_i = \sum_i J_i =0$ at intermediate steps.
The resulting expression for $B_1$ reads
\be{B1P}
B_1 = 8\pi G \int {d^3q\over 2|q| (2\pi)^3} \sum_{i,j} {p_ip_j \over qp_i qp_j } [p_i\overrightarrow{J}_{j}+ p_j\overleftarrow{J}_i]q~,~ \ee
where the arrows on $ J_i , J_j$ indicate whether the derivative acts on ${\cal M}_{N}$ or on its complex conjugate.

Clearly the basic integral to compute is:
\be{B1q}
	I^{\mu}_{ij} =  \int {d^3q\over 2|q|(2\pi)^3}  {p_ip_j q^\mu \over qp_i qp_j }   =  \int \frac{d^4q}{(2\pi)^3} \delta_+(q^2)  {p_ip_j q^\mu \over qp_i qp_j }~;~ \,\,\,\delta_+(q^2) = \delta (q^2) \Theta(-q_0) \, .
\ee

Unfortunately, for a given pair $i,j$ this integral has collinear divergences (that cannot be cured by inserting a lower limit on the soft-graviton's energy).
This can be easily seen in the fact that $qp_i$ goes like $\theta_{qi}^2$ when the angle  $\theta_{qi}$ between $\vect{q}$ and $\vect{p_i}$ goes to zero, while phase space (in four dimensions) gives $d\theta_{qi} \theta_{qi}$.
This problem disappears after summing over $i$ and $j$ but introducing a cut-off (meaning in this case a mass for the hard quanta) is quite cumbersome. A better way to proceed
is to modify $I_{ij}^{\mu}$ in 
such a way as
to make it collinear safe for each $i,j$ and $\mu$. We thus make the replacement
\be{Iijz}
I_{ij}^{\mu}\rightarrow \tilde{I}_{ij}^{\mu}= \int \frac{d^4q}{(2\pi)^3} \delta_+(q^2) \frac{\left[(p_{i}p_{j})q^{\mu}-(qp_{j})p_{i}^{\mu}-(qp_{i})p_{j}^{\mu}\right]}{(p_{i}q)(p_{j}q)}]\, .
\ee
The numerator now vanishes if $q$
is parallel to either 
$p_{i}$ or $p_{j}$,
removing the collinear singularities of each term.
Furthermore, one can check that the additional terms  drop out after summing over $i$ and $j$ thanks to linear and angular momentum conservation.

 At this point the calculation can be performed in two ways: by splitting the three-dimensional integral and  carrying out explicitly the angular integration, or by introducing an extra Lorentz invariant $\delta$-function to fix the GW frequency in a conveniently chosen Lorentz frame. In the first approach manifest Lorentz invariance is lost while in the second it is kept. We have checked that both approaches lead to the same final result.
Hereafter we follow the second, more elegant method.

 \subsection{Covariant calculation of $B_1$}

In order to arrive at a Lorentz-invariant 
 frequency spectrum let us define the following Lorentz covariant four-vector integral:
 \be{LCI}
K^{\mu}_{ij}(P,\Lambda) = \int \frac{d^4q}{(2\pi)^3} \frac{ \delta_+(q^2) \delta((qP/\Lambda^2) + 1)}{  (qp_i )(qp_j) }\left[(p_{i}p_{j})q-(qp_{j})p_{i}-(qp_{i})p_{j}\right]^{\mu} \, , 
\ee 
where $P$ is, a priori, an arbitrary four vector and $\Lambda$ is a constant with dimension of energy. Clearly $K^{\mu}_{ij}$ has nice transformation properties under Lorentz. 
 A physically interesting choice, adopted hereafter, consists in identifying  $P$  with the total momentum of the $n$ incoming (or $m$ outgoing) particles in a generic $n \rightarrow m$ process.
 In this case $\Lambda$ takes the meaning of $\sqrt{s} \hbar \omega_0$ where $s = -P^2$ is the Mandelstam variable of the corresponding channel and $ \omega_0$ is the center-of-mass frequency at which we wish to compute the spectrum.
The Lorentz-invariant (graviton number) spectrum  $\frac{dB_{1}}{d \omega_0}$ is then given in terms of an integral of the type (\ref{LCI}), i.e.:
 
 \beqn{dBLI}
\frac{dB_{1}}{d \omega_0} & =& \frac{8\pi G}{\hbar} \sum_{ij} \int \frac{d^4q}{(2\pi)^3} \frac{ \delta_+(q^2) \delta(\omega - \omega_0)}{  (qp_i )(qp_j) } [p_i\overrightarrow{J}_{j}+ p_j\overleftarrow{J}_i] \left[(p_{i}p_{j})q-(qp_{j})p_{i}-(qp_{i})p_{j}\right] \nonumber \\
&=& \frac{8\pi G}{\hbar} \sum_{ij} K^{\mu}_{ij}(P,  \omega_0)  [p_i\overrightarrow{J}_{j}+ p_j\overleftarrow{J}_i]_{\mu} \, ,
\eeqn
from which one can compute the GW energy spectrum upon multiplication by $\hbar  \omega_0$.

Lorentz covariance allows us to expand $K^{\mu}_{ij}$ in the form:
\be{LCexp}
K^{\mu}_{ij} = K_P P^{\mu} + K_i p_i^{\mu} + K_j p_j^{\mu} \, ,
\ee
where $K_P,  K_i,  K_j$ are functions of the four Lorentz-invariants $P^2 = - s, (Pp_i), (Pp_j)$ and $(p_ip_j)$. By its definition (\ref{LCI}) $K^{\mu}_{ij}$ is orthogonal to both $p_i^{\mu}$ and $p_j^{\mu}$ and therefore must be of the form:
\be{LCexp1}
K^{\mu}_{ij} = K \left[ (p_ip_j) P^{\mu} -  (Pp_j) p_i^{\mu} -  (Pp_i) p_j^{\mu}\right] \equiv K (p_ip_j)  {{Q}}_{ij}^{\mu}\, . 
\ee
Contracting $K^{\mu}_{ij} $ with $P^{\mu}$ and using the already known integral appearing in (\ref{Bagain}):
\be{integral1}
  \int {d^3q\over |q| } \delta(\omega - \omega_0) {p_i p_j \over qp_i qp_j } = - \frac{4 \pi}{\omega_0} \log \left(-{2p_i p_j\over m_i m_j}\right) 
\ee
as well as 
\be{integral2}
 \int {d^3q\over |q| } \delta(\omega - \omega_0) \left({Pp_i\over qp_i  } + {Pp_j\over qp_j  }\right) = \frac{4 \pi}{\omega_0} \log \left(-{4 Pp_i Pp_j\over P^2 m_i m_j}\right) \, ,
\ee
determines the scalar $K$ to be:
\be{J}
K = -  \frac{\omega_0\sqrt{s}}{4 \pi ^2  \tilde{s}_{ij} } \log\left[ \frac{-s (p_ip_j)}{2(Pp_i)(Pp_j)}\right] \, ,
\ee
where, also for future use, we have introduced the convenient quantity:
\be{stilde}
\tilde{s}_{ij} = -  {{Q}}_{ij}^2 =  s + \frac{ 2(Pp_i)(Pp_j)}{p_ip_j} \, .
\ee
Note the absence of singularities in (\ref{J}) when $\tilde{s}_{ij}$ vanishes.
Inserting this result in (\ref{dBLI}) we get (after renaming $\omega_0$ as $\omega$):
 \be{dBLI1}
\frac{dB_{1}}{d \omega} 
= - 2\frac{G\sqrt{s}}{\pi}  \sum_{ij}  \frac{1}{\tilde{s}_{ij} }  \log\left[ \frac{-s (p_ip_j)}{2(Pp_i)(Pp_j)}\right] {{Q}}_{ij}^{\mu}  [p_i\overrightarrow{J}_{j}+ p_j\overleftarrow{J}_i]_{\mu} \, .
\ee

 Let us elaborate further the result (\ref{dBLI1}) in the general case. Since the vector  ${{Q}}_{ij}^{\mu}$
  is orthogonal to both $p_i$ and $p_j$ we can replace $  [p_i\overrightarrow{J}_{j}+ p_j\overleftarrow{J}_i]_{\mu}$ with $(p_ip_j) (\overleftarrow{\frac{\partial}{\partial p_i}} + \overrightarrow{\frac{\partial}{\partial p_j}})_{\mu}$ and obtain, after multiplying by $\hbar \omega$, 
  \be{simpledB1}
  \frac{dE_{1}}{d \omega} 
= - 2 \frac{G\sqrt{s} \hbar \omega}{\pi}  \sum_{ij}   \frac{(p_ip_j)  }{ \tilde{s}_{ij} }  \log\left[ \frac{-s (p_ip_j)}{2(Pp_i)(Pp_j)}\right] {{Q}}_{ij}^{\mu} \Big(\overleftarrow{\frac{\partial}{\partial p_i}} + \overrightarrow{\frac{\partial}{\partial p_j}}\Big)_{\mu}\, . 
  \ee
 
 We see no reason for why this expression should vanish for a generic $n \rightarrow m$ process and explicit calculations seem to confirm this.
 We will see however that (\ref{simpledB1}) leads to a vanishing result in the case of a two-body collision. We will show this in two different ways: by working in a special Lorentz frame (the so-called Breit frame), and by using an appropriate Lorentz-covariant definition of the derivatives.
 
\subsection{Breit-frame argument for $B_1 =0$ in two-body scattering} 

In order to satisfy all kinematic constraints one can choose the Breit frame (BF) where
\be{BF}
p_1 = (E,p,k,0)
\quad p_2=  (E,-p,-k,0)
\quad p_3=  (-E,p,-k,0)
\quad p_4= (-E,-p,k,0)\, , 
\ee
with $E=\sqrt{p^2+k^2}$. 
In this frame the Mandelstam invariants read
\be{MvBF}
s = -(p_1+p_2)^2 = 4(p^2+k^2) \quad t= - (p_1+p_4)^2 = - 4 k^2 \quad u= - (p_1+p_3)^2 = - 4 p^2
\ee 
and obviously satisfy $ s+t+u = 0$. The only subtlety that we have to deal with is how to define $d/dp_{iz}$ in order to preserve $\vect{p}\cdot\vect{k} = 0$, where $\vect{p}= (p,0,0)$ is the `longitudinal' 3-momentum and $\vect{k}= (0,k,0)$ is the `transverse' 3-momentum. For $B_1$ this will turn out to be irrelevant since $d/dp_{i,z}$ combine with $p_{i,z}=0$ in the BF.  
 
Neglecting an overall factor, it is convenient to rewrite ${dB_1^{(i,j)}/d\omega}$ as
 \be{dB1}
 {dB_1^{(i,j)}\over d\omega} = \eta_i \eta_j E {1-\vect{v}_i\vect{v}_j \over 
 1+\vect{v}_i\vect{v}_j} \log{1-\vect{v}_i\vect{v}_j \over 2}  [ \delta^\mu_0 (1+\vect{v}_i\vect{v}_j) + \delta^\mu_r (\vect{v}_i+\vect{v}_j)^r] 
 (\overleftarrow{\partial}_{i,\mu} + \overrightarrow{\partial}_{j,\mu})]\, ,
 \ee
 where $\eta_i=+1(-1)$ for incoming (outgoing) particles. Using $\vect{v}_i = \vect{p}_i/E_i = \eta_i \vect{p}_i/E$ one finds
\be{vvv}
 \vect{v}_1 = {1\over E} (p,k,0) = - \vect{v}_2 \quad {\rm and}  \quad \vect{v}_3 = {1\over E} (-p,k,0) = - \vect{v}_4 \, . 
 \ee
 The relevant combinations are
 \beqn{v12}
&  \vect{v}_1 + \vect{v}_2 = 0 = \vect{v}_3 + \vect{v}_4 \quad , \quad   \vect{v}_1 + \vect{v}_3 = (0,{2k\over E},0)  = - \vect{v}_2 - \vect{v}_4 \, ,\nonumber \\
& \vect{v}_1 + \vect{v}_4  = ({2p\over E},0,0)  = - \vect{v}_2 - \vect{v}_3 \, , \nonumber \\
 & 1-\vect{v}_1\vect{v}_2 = 1-\vect{v}_3\vect{v}_4= 2 \quad, \quad 1-\vect{v}_1\vect{v}_3 = 1-\vect{v}_2\vect{v}_4 = {2p^2\over E^2} \, ,\nonumber \\
  & 1 +\vect{v}_1\vect{v}_2 = 1+\vect{v}_3\vect{v}_4= 0 \quad, \quad 1+\vect{v}_1\vect{v}_3 = 1+\vect{v}_2\vect{v}_4 = {2k^2\over E^2} \, , \nonumber \\
&1+\vect{v}_1\vect{v}_4 = 1+\vect{v}_2\vect{v}_3 = {2p^2\over E^2} \, . 
 \eeqn
 Combining the various terms one finds (${\rm reg}$ indicates regularisation by  a small mass)
\beqn{ff}
 &{dB_1^{(1,2)}\over d\omega}_{{\rm reg}} = E {1-\vect{v}_1\vect{v}_2 \over 
 1+\vect{v}_1\vect{v}_2} \log{1-\vect{v}_1\vect{v}_2 \over 2}  [ \delta^\mu_0 (1+\vect{v}_1\vect{v}_2) + \delta^\mu_r (\vect{v}_1+\vect{v}_2)^r] 
 (\overleftarrow{\partial}_{1,\mu} + \overrightarrow{\partial}_{2,\mu})]_{{\rm reg}} = 0 \, , \nonumber \\
 &{dB_1^{(1,3)}\over d\omega} = - E {1-\vect{v}_1\vect{v}_3 \over 
 1+\vect{v}_1\vect{v}_3} \log{1-\vect{v}_1\vect{v}_3 \over 2}  [ \delta^\mu_0 (1+\vect{v}_1\vect{v}_3) + \delta^\mu_r (\vect{v}_1+\vect{v}_3)^r] 
 (\overleftarrow{\partial}_{1,\mu} + \overrightarrow{\partial}_{3,\mu})] \nonumber \\
 &=- E {p^2\over k^2} \log{p^2\over E^2} [ \delta^\mu_0 {2k^2\over E^2}+ \delta^\mu_r (0,{2k\over E},0)^r] 
 (\overleftarrow{\partial}_{1,\mu} + \overrightarrow{\partial}_{3,\mu})] \nonumber \\ 
 & = -2 {p^2\over E^2} \log{p^2\over E^2} 
 [E \overleftarrow{\partial}_{E} + {E^2\over k} \overleftarrow{\partial}_{k} - E \overrightarrow{\partial}_{E} - {E^2\over k} \overrightarrow{\partial}_{k})]\, , \nonumber \\ 
 &{dB_1^{(1,4)}\over d\omega} = - E {1-\vect{v}_1\vect{v}_4 \over 
 1+\vect{v}_1\vect{v}_3} \log{1-\vect{v}_1\vect{v}_4 \over 2}  [ \delta^\mu_0 (1+\vect{v}_1\vect{v}_4) + \delta^\mu_r (\vect{v}_1+\vect{v}_4)^r] 
 (\overleftarrow{\partial}_{1,\mu} + \overrightarrow{\partial}_{4,\mu})]  \nonumber \\
&= - E {k^2\over p^2} \log{k^2\over E^2} [ \delta^\mu_0 {2p^2\over E^2}+ \delta^\mu_r ({2p\over E},0,0)^r] 
 (\overleftarrow{\partial}_{1,\mu} + \overrightarrow{\partial}_{4,\mu}] \nonumber \\
 &= -2 {k^2\over E^2} \log{k^2\over E^2} 
 [E \overleftarrow{\partial}_{E} + {E^2\over p} \overleftarrow{\partial}_{p} - E \overrightarrow{\partial}_{E} - {E^2\over p} \overrightarrow{\partial}_{p})]\, . 
\eeqn
 An interesting feature of the above result is that only combinations of derivatives of the kind $(k \overleftarrow{\partial}_{E} + E\overleftarrow{\partial}_{k})$ or 
 $(p \overleftarrow{\partial}_{E} + E\overleftarrow{\partial}_{p})$ appear. Since these combinations vanish when applied to the constraint $E^2 - p^2 - k^2 =0$ their action does not depend on which independent variables we use to express the amplitude.
 
We now note that $dB_1^{(3,1)}$ is obtained form $dB_1^{(1,3)}$ by simply exchanging the sense of the arrows which amounts to simply changing an overall sign.
Therefore $dB_1^{(3,1)}+ dB_1^{(1,3)} =0$ and similarly for $dB_1^{(1,4)}$ and $dB_1^{(4,1)}$.
In conclusion one gets 
\be{hjk}
 {dB_1^{BF}\over d\omega} = \frac{d E_1^{GW}}{d \omega} = 0\, . 
 \ee

\subsection{Covariant argument for $B_1=0$   in two-body scattering}

In the case of an elastic $2 \rightarrow 2$ process things simplify. Recalling that now $P = p_1 + p_2 = -p_3 - p_4$,
let us first notice that  the contributions with $i=j$ vanish trivially. Also the contributions
with   $i,j = 1,2$ and $i,j = 3,4$ vanish since the vectors ${{Q}}_{12}^{\mu}, {{Q}}_{34}^{\mu}$ vanish identically.

We are thus led to consider the remaining pairs: $i,j = 1,3$, $i,j = 2,4$ and $i,j = 2,3$, $i,j = 1,4$. These need special attention since it is not a priori obvious how
the partial derivatives are defined since the $N(=4)$ momenta are constrained by momentum conservation $\sum p_i = 0$ and by the mass-shell conditions $p_i^2 =0$.
In terms of the Mandelstam variables $s,t,u$ of the four-point amplitude the constraint $s+t+u = 0$ holds.

By studying carefully how the soft theorems work at the level of the five-point function (see e.g. the one in Appendix C) with a soft, but finite momentum graviton, one can argue 
 that the correct way to define the derivatives is to first replace the Mandelstam variables as follows:
\beqn{defDeltas}
s &\rightarrow& - \Delta_s^2~;~ \,\,\,\Delta_s = \frac12(p_1+p_2-p_3-p_4) \nonumber \\
t &\rightarrow& - \Delta_t^2~;~ \,\,\,\Delta_t = \frac12(p_1+p_4-p_2-p_3) \nonumber \\
u &\rightarrow& - \Delta_u^2~;~ \,\,\,\Delta_u = \frac12(p_1+p_3-p_2-p_4) 
\eeqn
and by then letting the derivatives act on the modified Mandelstam variables as if all the momenta were independent (in a sense they are, because in the five-point function $\sum p_i = - q$). 
With these rules the basic derivatives become:
\beqn{deriv}
\partial_1^{\mu} &=& -  \Delta_s^{\mu} \partial_s - \Delta_t^{\mu} \partial_t - \Delta_u^{\mu} \partial_u \nonumber \\
\partial_2^{\mu} &=& -  \Delta_s^{\mu} \partial_s + \Delta_t^{\mu} \partial_t + \Delta_u^{\mu} \partial_u \nonumber \\
\partial_3^{\mu} &=& + \Delta_s^{\mu} \partial_s  + \Delta_t^{\mu} \partial_t - \Delta_u^{\mu} \partial_u \nonumber \\
\partial_4^{\mu} &=& + \Delta_s^{\mu} \partial_s  - \Delta_t^{\mu} \partial_t +\Delta_u^{\mu} \partial_u \;.
\eeqn
As a consequence one can easily check that:
\be{sumder}
p_1 \partial_1 = p_2 \partial_2 = p_3 \partial_3 = p_4 \partial_4 = \frac12 (s \partial_s +t \partial_t +u \partial_u) ~;~ \sum_{i=1}^4 p_i \partial_i = 2 (s \partial_s +t \partial_t +u \partial_u)
\ee
as one would naively expect. The above rules are also consistent with angular momentum conservation (which we have used already). Indeed:
\be{J1}
J_1^{\mu \nu} = p_1^{\mu}  \partial_1^{\nu} - (\mu \leftrightarrow \nu)  = - p_1^{\mu}  (\Delta_s^{\nu} \partial_s + \Delta_t^{\nu} \partial_t + \Delta_u^{\nu} \partial_u)  - (\mu \leftrightarrow \nu)
\ee
and, after adding to (\ref{J1}) $J_2, J_3$ and $J_4$, one easily finds:
\be{Jcons}
\sum_{i=1}^4 J_i^{\mu \nu} =  - \Delta_s^{\mu}  \Delta_s^{\nu} \partial_s - \Delta_t^{\mu}  \Delta_t^{\nu} \partial_t - \Delta_u^{\mu} \Delta_u^{\nu} \partial_u - (\mu \leftrightarrow \nu) = 0
\ee
These partial-derivative rules will be taken up again when discussing the ${{\cal O}}(\omega^2)$ correction.

For instance, consider the contribution from $(i,j) = (1,3)$ to which we must add, of course, the one from $(i,j) = (3,1)$. As a result, the sum $\overrightarrow{\partial}_{1} + \overrightarrow{\partial}_{3}$ appears. Because of the relative signs with which $p_1$ and $p_3$ appear in (\ref{deriv}) only the term $\Delta_u^{\mu} \partial_u$ will survive. However, ${{Q}}_{13}{\cdot}\Delta_u =0$ from the transversality of ${{Q}}_{13} $.
Very similar arguments show that all contributions to $B_1$ vanish identically\footnote{At least if we neglect  log corrections that appear to sub-leading order beyond tree-level. On the other hand, recent studies \cite{Sahoo:2018lxl}, \cite{CCV18} that keep those effects into account appear to confirm this conclusion.} in agreement with the Breit-frame conclusion (\ref{hjk}).

We see however no reason for the same trivial result to hold for $N \ge 5$. Also, as we shall see in the following Section, a non-vanishing result will emerge at the next order in $\omega$.
\section{The ${{\cal O}}(\omega^2)$ correction for a generic massless process}
\setcounter{equation}{0}

At tree level, the sub-sub-leading terms come from the sum of three contributions:
\be{B2}
B_{2}=B_{02}+B_{20}+B_{11}=8\pi G\int \frac{d^{3}q}{2(2\pi)^{3}|q|}\sum_{s} \sum_{ij}[S_{0}S_{2}^{*}+
S_{2}S_{0}^{*}+S_{1}S_{1}^{*}]_{ij}\, ,
\ee
where $s$ stands for the graviton's polarization and
\be{T02}
[S_{0}S_{2}^{*}]_{ij}=- {\cal S}_{0}\frac{p_{i}hp_{i}}{qp_i}\frac{q\overrightarrow{J}_{j}h^{*}\overrightarrow{J}_{j}q}{2 qp_{j}} {\cal S}_{0}^{*}~~  ; ~~[S_{2}S_{0}^{*}]_{ij}=-{\cal S}_{0}\frac{q\overleftarrow{J}_{i}h\overleftarrow{J}_{i}q}{2 qp_{i}} \frac{p_{j}hp_{j}}{qp_j}{\cal S}_{0}^{*}
\ee

\be{T11}
[S_{1}S_{1}^{*}]_{ij}= {\cal S}_{0}\frac{p_{i}h\overleftarrow{J}_{i}q}{qp_{i}}\frac{p_{j}h^{*}\overrightarrow{J}_{j}q}{qp_{j}} {\cal S}_{0}^{*}\, .
\ee
As in the previous Section we have introduced arrows to indicate which amplitude (${\cal S}_0$ or ${\cal S}_0^*$) the operators $J_i, J_j$ act on.
However, unlike in the sub-leading case, we are now facing the problem of how to order the double derivatives appearing in $S_2$. 
Hereafter we will use the prescription\footnote{We acknowledge useful correspondence with Matin Mojaza and Paolo Di Vecchia about this important issue.} that all derivatives act on ${\cal S}_0$ or ${\cal S}_0^*$ {\it before} any possible multiplying factor. We will come back to this possible ambiguity when discussing specific processes. This being said, in the following we shall usually omit to write  ${\cal S}_0, {\cal S}_0^*$ in the formulae.
Summing over  $s=\pm 2$, one gets
\be{Wdef1}
[S_{0}S_{2}^{*}+
S_{2}S_{0}^{*}+S_{1}S_{1}^{*}]_{ij} = \frac{q^{\mu'} q^{\nu'} {\cal W}^{ij}_{\mu'\nu'}}{qp_{i} qp_{j}}\ ,
\ee
where:
\be{Tmun}
{\cal W}^{ij}_{\mu'\nu'}=\Pi_{\mu\nu,\rho\sigma}\Big\{
-{1\over 2} [p_{i}^{\rho}p_{i}^{\sigma}\overrightarrow{J}_{j\mu'}{}^{\mu}\overrightarrow{J}_{j\nu'}^{\nu}+(i\leftrightarrow j)] + 
p_{i}^{\mu}\overleftarrow{J}^{\nu}_{i\mu'}p^{\rho}_{j}\overrightarrow{J}_{j\nu'}^{\sigma} \Big\} \, ,
\ee
and $\Pi_{\mu\nu,\rho\sigma}$ was already defined in (\ref{tttensor}).

Thanks to gauge invariance, momentum conservation and Lorentz invariance, we expect that all terms depending on $\bar{q}$ vanish after summing over $i$ and $j$. This is easily seen if one observes that $\bar{q}$ always appears in combination with $q$ in $\Pi$. Since $qJ_iq= 0= qJ_jq$, $q$ will have to contract either with $p_i$ or with $p_j$ (or with both). In any case the factor $qp_i$ (or $qp_j$) will cancel the pole and the sum over $i$ (or $j$) will vanish thanks to momentum or angular momentum conservation.
As a result, like in previous cases, we can replace ${\pi}_{\mu\nu}$ with $\eta_{\mu\nu}$ and $\Pi_{\mu\nu,\rho\sigma}$ with 
${1\over 2} (\eta_{\mu\rho} \eta_{\nu\sigma} + \eta_{\mu\sigma} \eta_{\nu\rho}-\eta_{\mu\nu} \eta_{\rho\sigma})$.
The final result reads
\beqn{Wdef}
2 {\cal W}_{\mu\nu}^{ij}  &=& -[(\overrightarrow{J}_jp_i)_\mu (p_i\overrightarrow{J}_j)_\nu + (\overleftarrow{J}_ip_j)_\mu (p_j\overleftarrow{J}_i)_\nu] \nonumber \\
&+& (\overleftarrow{J}_ip_i)_{(\mu} (p_j\overrightarrow{J}_j)_{\nu)} 
- (p_ip_j) (\overleftarrow{J}_i\overrightarrow{J}_j)_{(\mu\nu)} - (\overrightarrow{J}_jp_i)_{(\mu} (p_j\overleftarrow{J}_i)_{\nu)} \, , \eeqn
where $(\mu  \nu)$ means symmetrization with strength one. Hereafter it will be understood that $J_i, J_j$ carry a left, resp. right, arrow.

In conclusion, $B_{2}=B_{02}+B_{20}+B_{11}$ takes the form 
\be{B2qqW}
B_2 = {G\over 2\pi^2} \int {d^3q\over |q|} \sum_{i,j} {q^\mu q^\nu \over qp_i qp_j} {\cal W}_{\mu\nu}^{ij} =
{G\over 2\pi^2} \sum_{i,j} M^{\mu\nu}_{ij} {\cal W}_{\mu\nu}^{ij} \, ,
\ee
where ${\cal W}_{\mu\nu}^{ij}$ is given in (\ref{Tmun}) and
\be{B22}
M^{\mu\nu}_{ij} =  \int {d^3q \over |q|} \frac{q^\mu q^\nu }{qp_{i}qp_{j}}\, .
\ee
The contraction ${\cal W}_{\mu\nu}^{ij} q_{\mu} q_{\nu}$ gives the following terms (from $B_{02}+B_{20}$ and $B_{11}$, respectively):
\beqn{Wcontractions}
&&\left.\frac{{\cal W}_{\mu\nu}^{ij} q_{\mu} q_{\nu}}{(qp_{i})(qp_{j})}\right\vert_{20+02} = -
\sum_{i,j}\left[\frac{(q\overleftarrow{J}_{i}p_{j})(p_j\overleftarrow{J}_{i}q)+(q\overrightarrow{J}_{j}p_{i})(p_i\overrightarrow{J}_{j}q)}{2(qp_{i})(qp_{j})} \right] \\
&&\left.\frac{{\cal W}_{\mu\nu}^{ij} q_{\mu} q_{\nu}}{(qp_{i})(qp_{j})}\right\vert_{11}=\sum_{i,j}\frac{(p_{j}\overleftarrow{J}_{i}q)(p_{i}\overrightarrow{J}_{j}q){-}(p_{i}\overleftarrow{J}_{i}q)(p_{j}\overrightarrow{J}_{j}q){-}(p_{i}p_{j})(q\overleftarrow{J}_{i}\overrightarrow{J}_{j}q)}{2(qp_{i})(qp_{j})}.
\eeqn

As it was already the case for $B_1$, the individual integrals for fixed $i$ and $j$ are affected by `collinear' divergences, when $\vect{n}=\vect{q}/\omega$ is  parallel to either $\vect{v}_i$ or $\vect{v}_j$. Although these divergences cancel after summing over $i$ and $j$, it is convenient to shift the integration variable $q$ in such a way that integrals be finite for fixed $i$ and $j$. 
The obvious choice for the shift is the one already used for $B_1$.  

That shift is ill defined for $i=j$, but there were no such contributions in $B_1$. Terms with $i=j$ are present in  $B_2$ and have to be treated separately.
Instead, for $i\ne j$, we apply the replacement
\be{shiftdef} 
q = \tilde{q}_{ij} + \frac{qp_j}{p_ip_j}p_i  + \frac{qp_i}{p_ip_j}p_j \, .
\ee
 In the following, to simplify notation, we will often suppress the indices $ij$ in $\tilde{q}_{ij} $. Notice that while $q^2 = 0$, $\tilde{q}^2 = -2 {qp_iqp_j/p_ip_j}$

Also at variance with what happened for $B_1$, shifting $q$ does not leave individual integrals appearing in $B_2$ unchanged. The explicit calculation is reported in Appendix A. The final result of such calculation is:
\beqn{B2new}
B_2 = {G\over 2\pi^2} \int {d^3q\over |q|} \sum_{i,j} {q^\mu q^\nu \over qp_i qp_j} {\cal W}_{\mu\nu}^{ij} = {G\over 2\pi^2} \Big\{\int {d^3q\over |q|} \sum_{i\neq j} {\tilde{q}^\mu \tilde{q}^\nu \over qp_i qp_j} {\cal W}_{\mu\nu}^{ij} \nonumber\\
-  \Big[\frac{3}{2}\sum_i \overleftarrow{D}_{i}\sum_j\overrightarrow{D}_{j} - 2 \sum_i
(\overleftarrow{D}_{i} + \overrightarrow{D}_{i})^2)
\Big] \int {d^3q\over |q|}\Big\} \, ,
\eeqn
where $D_{i}=p_{i}\partial_{i}~~ ({\rm no ~ sum})$.  
We now proceed as in the case of $B_1$ by
introducing a Lorentz invariant  constraint such as $\delta((qP/\Lambda^2) + 1)$, with $P$ taken to be the total incoming momentum, $P=p_1+p_2$, and $\omega_0 = {\Lambda^2}{/}{\sqrt{s}}$ (where $s=- P^2 = 4E^2 = E_{CM}^2$) the center-of-mass frequency at which we wish to compute the gravitational wave spectrum.

The last integral  can be easily computed  in the CM frame where $P=(2E,\vect{0})$ and $q=-\omega(1, \vect{n})$ (emitted radiation, $\eta=-1$) 
\be{phsp}
\int {d^3q\over |q|} \delta\Big({qP\over \Lambda^2} + 1\Big) = \int \omega d\omega d\Omega_{\vect{n}} \delta\Big(1 -{ \omega\over\omega_0} \Big) =  4\pi \omega_0^2 \, .
\ee

In order to evaluate the remaining `shifted' integrals:
\be{B2integrals}
\widetilde{M}_{ij}^{\mu\nu}= \int {d^3q\over |q|} \sum_{i,j} {\tilde{q}^\mu \tilde{q}^\nu \over qp_i qp_j} \delta	\Big({qP\over \Lambda^2} + 1\Big) \, ,
\ee
one observes that $\widetilde{M}_{ij}^{\mu\nu} = \widetilde{M}_{ij}^{\nu\mu}$ (symmetric) and
$p_{i,\mu}\widetilde{M}_{ij}^{\mu\nu}  = 0 = p_{j,\mu}\widetilde{M}_{ij}^{\mu\nu}$ (bi-transverse).
As a consequence $\widetilde{M}_{ij}^{\mu\nu}$ can be written in the form 
$$
\widetilde{M}_{ij}^{\mu\nu} = A P_{ij}^{\mu\nu} + B H_{ij}^{\mu\nu} \, ,
$$ 
where
$$
P_{ij}^{\mu\nu} =\left(P^\mu - {Pp_j\over p_ip_j} p_i^\mu  - {Pp_i\over p_ip_j} p_j^\mu \right)\left(P^\nu - {Pp_j\over p_ip_j} p_i^\nu  - {Pp_i\over p_ip_j} p_j^\nu \right) 
= {{Q}}_{ij}^{\mu} {{Q}}_{ij}^{\nu}
$$ 
and
$$
H_{ij}^{\mu\nu} = \eta^{\mu\nu}  - {p_i^\mu  p_j^\nu + p_i^\nu  p_j^\mu\over p_ip_j}  \, .
$$

After `tracing' with $\eta_{\mu\nu}$ and contracting with $P_\mu P_\nu$ one gets
\be{ABdef}
A = {\eta_{\mu\nu} \widetilde{M}_{ij}^{\mu\nu} \over \tilde{s}_{ij}}  + 2  {P_{\mu} P_\nu \widetilde{M}_{ij}^{\mu\nu} \over \tilde{s}^2_{ij}} \qquad , \qquad B = {\eta_{\mu\nu} \widetilde{M}_{ij}^{\mu\nu}}  +  {P_{\mu} P_\nu \widetilde{M}_{ij}^{\mu\nu} \over \tilde{s}_{ij}} \, ,
\ee
where we encounter again:
\be{stildeij}
\tilde{s}_{ij} = -{{Q}}^2_{ij} = - P^2 + 2 {Pp_i Pp_j\over p_ip_j} = s + 2 {Pp_i Pp_j\over p_ip_j} \, .
\ee

The first `scalar' integral $\eta_{\mu\nu} \widetilde{M}_{ij}^{\mu\nu}$ is easily computed
\be{etaM}
\eta_{\mu\nu} \widetilde{M}_{ij}^{\mu\nu} = \int {d^3q\over |q|} {\tilde{q}^2 \over qp_i qp_j} \delta\Big(1 -{ \omega\over\omega_0} \Big) = - {2\over p_i p_j} \int {d^3q\over |q|} \delta\Big(1 -{ \omega\over\omega_0} \Big)  = - {2\over p_i p_j}
4\pi\omega_0^2 \, .
\ee
while the second `scalar' integral $ P_{\mu}P_{\nu} \widetilde{M}_{ij}^{\mu\nu}$ requires more work and gives the result (see Appendix B, Eq. (\ref{PMPfin1})): 
\be{PMPfin}
P_{\mu}P_{\nu} \widetilde{M}_{ij}^{\mu\nu} =  {8\pi  \omega_0^2 \over p_i p_j}
\left\{ {P^2\over 2} \log {P^2 p_i p_j \over 2 Pp_i Pp_j} + \tilde{s}_{ij} \right\}\, .
\ee

Plugging this and $\eta_{\mu\nu} \widetilde{M}_{ij}^{\mu\nu} = - {8\pi P^2 \omega_0^2/P^2 p_i p_j} = -8\pi \omega_0^2 / p_i p_j$
in (\ref{ABdef})
yields 
\be{Afinal}
A = {8\pi \omega_0^2 \over p_i p_j \tilde{s}_{ij}} \left(1  + { P^2 \over \tilde{s}_{ij}} \log {P^2 p_i p_j \over 2 Pp_i Pp_j} \right)
\ee
and
\be{Bfinal}
B =  {4\pi \omega_0^2 P^2\over p_i p_j \tilde{s}_{ij}} \log {P^2 p_i p_j \over 2 Pp_i Pp_j}\, . \ee

Let us now combine the integral $\widetilde{M}_{ij}^{\mu\nu} =  A P_{ij}^{\mu\nu} + B H_{ij}^{\mu\nu}$
with the tensor ${\cal W}_{\mu\nu}^{ij}$ of (\ref{Tmun}). Using bi-trasversality of $H$ and $P$, many terms drop from ${\cal W}_{\mu\nu}^{ij}$ upon contraction, {\it viz.} 
\be{Wijarrow}
2 {\cal W}_{\mu\nu}^{ij} \rightarrow
 (p_ip_j)^2 (\overleftarrow{\de}_{i\mu} \overleftarrow{\de}_{i\nu} + \overrightarrow{\de}_{j\mu} \overrightarrow{\de}_{j\nu}) 
+(p_ip_j)^2 (\overleftarrow{\de}_{i\mu} \overrightarrow{\de}_{j\nu}+\overleftarrow{\de}_{i\nu} \overrightarrow{\de}_{j\mu}) 
 \ee
 $$=(p_ip_j)^2 (\overleftarrow{\de}_{i\mu}+\overrightarrow{\de}_{j\mu})
 (\overleftarrow{\de}_{i\nu} + \overrightarrow{\de}_{j\nu}) \, .$$

Contracting with $(B)H$ and $(A)P$ yields 
\be{HW}
-2 H^{\mu\nu}_{ij} {\cal W}_{\mu\nu}^{ij} = - (p_ip_j)^2 (\overleftrightarrow{\de_{ij}})^{2} + 2(p_ip_j) p_i \overleftrightarrow{\de_{ij}} p_j \overleftrightarrow{\de_{ij}} \, ,
\ee
\be{PW}
-2P^{\mu\nu}_{ij} {\cal W}_{\mu\nu}^{ij}=  - (p_ip_j)^2 \left( \Pi^{\mu}  ( \overleftrightarrow{\de_{ij}})_{\mu}\right)^2  \, ,
\ee
where $(\overleftrightarrow{\de_{ij}})_{\nu}\equiv (\overleftarrow{\de})_{i\nu} + (\overrightarrow{\de})_{j\nu}$.
Including the coefficient functions $A$ and $B$, defined in (\ref{Afinal}) and (\ref{Bfinal}), and inserting the result in Eqs. (\ref{B2new}, \ref{B2integrals}), we arrive at our final expression for $B_2$ (after renaming again $\omega_0$ as $\omega$):
\beqn{B2fin2}
B_2 |{\mathcal S}_{if}|^2 &=&    {\mathcal S}_{if}^{\dagger} \frac{G \omega^2}{\pi} \left( C_{1} +C_2 +   C_3 \right)  {\mathcal S}_{fi}  \nonumber \\
C_{1}&=& - 3 \sum_i \overleftarrow{D}_i \sum_j \overrightarrow{D}_j + 4 \sum_i (
\overleftarrow{D}_i + \overrightarrow{D}_i)^2
\nonumber \\
C_2 &=&    \sum_{i\neq j}{P^2\over  \tilde{s}_{ij}} \log {P^2 p_i p_j \over 2 Pp_i Pp_j} [p_ip_j (\overleftrightarrow{\de_{ij}})^2 - 2 p_i(\overleftrightarrow{\de_{ij}}) p_j(\overleftrightarrow{\de_{ij}}) ] \nonumber \\ 
C_3 &=& \sum_{i\neq j} {2 \over p_i p_j \tilde{s}_{ij}}  \left[ 1 + {P^2\over \tilde{s}_{ij}}  \log {P^2 p_i p_j \over 2 Pp_i Pp_j}\right]  (p_ip_j)^2 \left( {{Q}}_{ij}^{\mu}  (\overleftrightarrow{\de_{ij}})_{\mu}\right)^2 \, ,
\eeqn
where we recall that  ${{Q}}_{ij}^{\mu} \equiv \left(P^\mu - {Pp_j\over p_ip_j} p_i^\mu  - {Pp_i\over p_ip_j} p_j^\mu \right)$ and that, by convention, all derivatives act on the amplitude before any possible multiplier.


\section{Specializing to   two-body collision processes}
\setcounter{equation}{0}

\subsection{General covariant result for $2 \to 2$ scattering}

As in the case of $B_1$ a big simplification  occurs for a four-point amplitude. We will continue using the recipe for the derivatives that led to a vanishing result for $B_1$.

In the case of $C_1$ things are very simple. Using Eq.(\ref{sumder}) we get:
\beqn{C1new}
C_1 = - 3 \sum_i \overleftarrow{D}_i \sum_j \overrightarrow{D}_j + 4 \sum_i (
\overleftarrow{D}_i + \overrightarrow{D}_i)^2 = 12 \overleftarrow D \overrightarrow D+ 4  (
\overleftarrow{D}+ \overrightarrow{D})^2\, . 
\eeqn

Let us now consider $C_{2}^{ij}$: it can be re-expressed as follows: 
\be{C2ij}
C_2^{ij} = C_2^{ji} =  K_2^{ij} {{L}}_2^{ij} ~~,~~ 
\ee
where:
\be{Kij}
K_2^{12} = K_2^{34}  =  -1 \: , \quad  K_2^{13} = K_2^{24} =    \frac{u}{t} \log \left(- \frac{u}{s}  \right) \: , \quad K_2^{14} = K_2^{23} =  \frac{t}{u} \log \left(- \frac{t}{s}  \right)
\ee
and
\be{Oij}
{{L}}_2^{ij} + {{L}}_2^{ji} = H_{ij}^{\mu\nu}\left[ (  \overleftarrow{\de}_{i\mu} + \overrightarrow{\de}_{j\mu}) (  \overleftarrow{\de}_{i\nu} + \overrightarrow{\de}_{j\nu}) + (i \leftrightarrow j) \right]\, .
\ee
Let us consider, as an example, the case of ${{L}}_2^{12} + {{L}}_2^{21}$ and use our rules (\ref{deriv}) for taking derivatives. Then:
\be{O12}
{{L}}_2^{12} + {{L}}_2^{21} = \frac14 H^{12}_{\mu\nu} (p_3 - p_4)^{\mu} (p_3 - p_4)^{\nu}   2  \overleftrightarrow{\Delta}_{tu}^2 = - tu \overleftrightarrow{\Delta}_{tu}^2 \, ,
\ee
where we have used $ H_{12}^{\mu\nu} (p_3 - p_4)_{\mu} (p_3 - p_4)_{\nu} = 4H_{12}^{\mu\nu} (p_3)_{\mu} (p_3 )_{\nu} =  - 2 t u$ and
\beqn{Deltatu}
 &&( \overleftarrow{\de}_{tt} + \overleftarrow{\de}_{uu} - 2 \overleftarrow{\de}_{t}  \overleftarrow{\de}_{u} )  - 2 (\overleftarrow{\de}_{t} \overrightarrow{\de}_{t} + \overleftarrow{\de}_{u} \overrightarrow{\de}_{u} - \overleftarrow{\de}_{t}\overrightarrow{\de}_{u} - \overleftarrow{\de}_{u}\overrightarrow{\de}_{t})  \nonumber \\ 
 &&  + (\overrightarrow{\de}_{tt} +\overrightarrow{\de}_{uu}  - 2 \overrightarrow{\de}_{t}  \overrightarrow{\de}_{u})   =  (\overleftarrow{\de}_{t}- \overrightarrow{\de}_{t} - \overleftarrow{\de}_{u} + \overrightarrow{\de}_{u})^2 \equiv  \overleftrightarrow{\Delta}_{tu}^2 \, .
\eeqn

We have also used the orthogonality between $H_{12}$ and $p_1, p_2$ to get rid of the derivatives w.r.t. $s$. The same result is obtained for ${{L}}_2^{34} + {{L}}_2^{43}$ and,
mutatis mutandis,  for the other pairs. Summing up the different pairs, each with its own $K_2^{ij}$ factor gives:
\be{C2}
C_2 = + 2tu \overleftrightarrow{\Delta}_{tu}^2  - 2st \log\left(- \frac{t}{s} \right) \overleftrightarrow{\Delta}_{su}^2 - 2su \log\left(- \frac{u}{s} \right) \overleftrightarrow{\Delta}_{st}^2  \, .
\ee

Consider finally $C_3$. In this case there is no contribution from $i,j = 1, 2$ and $i,j = 3,4$ since, as already noticed for $B_1$, ${{Q}}_{12} ={{Q}}_{34} = 0$.
Like for $C_2^{ij}$ we can decompose $C_3^{ij}$ as:
\be{C3ij}
C_3^{ij} = C_3^{ji} =  K_3^{ij} {{L}}_3^{ij} ~~;~~ \, ,
\ee
where:
\beqn{K3ij}
K_3^{13} = K_3^{24} =    \frac{4}{ts} \left[ 1 + \frac{u}{t} \log \left(- \frac{u}{s}  \right) \right] \quad , \quad 
K_3^{14} = K_3^{23} =  \frac{4}{us}  \left[ 1 + \frac{t}{u} \log \left(- \frac{t}{s}  \right) \right] \, ,
\eeqn
and
\be{O3ij}
{{L}}_3^{ij}  = \left((p_ip_j) {{Q}}_{ij}^{\mu} (  \overleftarrow{\de}_{i\mu} + \overrightarrow{\de}_{j\mu}) \right)^2 \, .
\ee
Consider, for instance, ${{L}}_3^{13}$.  Using the rules (\ref{deriv}) and taking into account the orthogonality between $\Pi_{13}$ and $p_1, p_3$, we easily find:
\beqn{O313}
&{{L}}_3^{13} = {{L}}_3^{31} = {{L}}_3^{24} = {{L}}_3^{42} = \left[(p_1p_3) ({{Q}}_{13} p_2)  (\overleftarrow{\de}_{s}{-}\overleftarrow{\de}_{t}{-}\overrightarrow{\de}_{s}{+}\overrightarrow{\de}_{t})\right]^2 = \frac{s^2t^2}{4} \overleftrightarrow{\Delta}_{st}^2,   \nonumber \\ 
&{{L}}_3^{14} = {{L}}_3^{41} = {{L}}_3^{23} = {{L}}_3^{32}=\left[(p_1p_4) ({{Q}}_{14} p_2)  (\overleftarrow{\de}_{s}{-}\overleftarrow{\de}_{u}{-}\overrightarrow{\de}_{s}{+}\overrightarrow{\de}_{u})\right]^2 =\frac{s^2u^2}{4} \overleftrightarrow{\Delta}_{su}^2,
\eeqn
and
\beqn{C3}
C_3 =   4st \overleftrightarrow{\Delta}_{st}^2  + 4su \overleftrightarrow{\Delta}_{su}^2 + 4st \log\left(- \frac{t}{s} \right) \overleftrightarrow{\Delta}_{su}^2 + 4su \log\left(- \frac{u}{s} \right) \overleftrightarrow{\Delta}_{st}^2  \, .
\eeqn

Adding up finally  all the contributions  we obtain: 
$$\sum_{i} C_1^{ii}+\sum_{i\neq j}C_{1}^{ij}+  \sum_{i,j} \left(  C_2^{ij} +   C_3^{ij} \right) 
= - 4 \overleftarrow{D}\overrightarrow{D} + 4( \overleftarrow{D}^{2} + \overrightarrow{D}^{2})  $$
\be{final} 
+2tu \overleftrightarrow{\Delta}_{tu}^2 +4st \overleftrightarrow{\Delta}_{st}^2+4su \overleftrightarrow{\Delta}_{su}^2 +2st \log\left(- \frac{t}{s} \right) \overleftrightarrow{\Delta}_{su}^2 
+2su \log\left(- \frac{u}{s} \right) \overleftrightarrow{\Delta}_{st}^2  \, , 
\ee
where we recall that:
\be{defs}
D \equiv s \de_s + t \de_t + u \de_u ~:~ \overleftrightarrow{\Delta}_{st} \equiv \left( \overleftarrow{\de}_{s} - \overleftarrow{\de}_{t}  - \overrightarrow{\de}_{s} + \overrightarrow{\de}_{t} \right)\, \dots
\ee
and 
\be{dE2dom}
B_2 |{\mathcal S}_{if}|^2 =  \frac{G \hbar \omega^2}{\pi}  {\mathcal S}_{if}^{\dagger} \left(\sum_{i}C_1^{ii} +\sum_{i\neq j}C_{1}^{ij}+  \sum_{ij} \left( C_2^{ij} +   C_3^{ij} \right) \right) {\mathcal S}_{fi} = \frac{dE_2^{GW}}{d (\hbar \omega)} | {\mathcal S}_{if}|^2 \, .
\ee
Note that the operators $D$ and $\overleftrightarrow{\Delta}_{st}, \dots$ are unambiguous in the sense that they give the same result when acting on $A(s,t)$, $\tilde{A}(s,u) \equiv A(s, -s-u)$ or  $\hat{A}(t,u) \equiv A(-u-t, t)$.

The previous expression can be further simplified by using the easily proven identity:
\be{identity}
[tu\overleftrightarrow{\Delta}_{tu}^2{+}st \overleftrightarrow{\Delta}_{st}^2{+}su \overleftrightarrow{\Delta}_{su}^2] = -(\overleftarrow{D}{-}\overrightarrow{D})^2 \, ,
\ee
allowing us to re-express  $B_{2}$ as follows: 
\beqn{fin4p}
&&B_{2}  |{\mathcal S}_{if}|^2  = \frac{dE_2^{GW}}{d (\hbar \omega)} | {\mathcal S}_{if}|^2 = 2\frac{G \hbar \omega^2}{\pi} \times \\
&& {\cal S}_{if}^{\dagger}  \left\{ \overleftarrow{D}^2 + \overrightarrow{D}^2 + \left[st + us\log\left(-{u\over s}\right)\right] \overleftrightarrow{\Delta}^2_{st} + \left[su + ts\log\left(-{t\over s}\right)\right] \overleftrightarrow{\Delta}^2_{su}\right\} {\cal S}_{fi}  \nonumber \, . 
\eeqn 
A few comments on (\ref{fin4p}) are in order:
\begin{itemize}
\item The result is symmetric, as it should, between $t$ and $u$. It is not, instead, with respect to $s$ since we are working in the $s$-channel center of mass;
\item The $(1,1)$ contribution to (\ref{fin4p}), being an absolute square, should be positive. This can be checked to be the case by isolating the terms that contain one derivative acting on ${\cal S}$ and one acting on ${\cal S}^{\dagger}$. Since the combination of operators appearing in $\overleftrightarrow{\Delta}^2_{st}$ and $\overleftrightarrow{\Delta}^2_{su}$ are negative-definite, their pre factors must be negative as well. It is straightforward to prove that this is indeed the case in the $s$-channel physical region.
\item A non trivial check of (\ref{fin4p}) consists in considering the $t \to 0$ (or equivalently, given its symmetry, the $u \to 0$) limit. If we go back to Eq. (\ref{softoper}) for a $2 \rightarrow 2$ process, and consider the forward limit $p_2+p_3, p_1+p_4 \to 0$, we can easily check that both $S_0$ and $S_2$ (at least naively) vanish in that limit while $S_1$ does not.
This is because, in the limit, the contributions of $i=2$ and $i=3$ have opposite denominators and equal (opposite) numerators for $S_0$ and $S_2$ ($S_1$). The same happens of course for the $i=1$ and $i=4$ contributions. This is confirmed by the vanishing of the leading term $S_0^* S_0$ in the forward ($t=0$) or backward ($u=0$) direction, see Eq. (\ref{ZFL4p}). 
Naively, one should expect the same to be the case for the $S_0^* S_2$ interference term. Such terms are easily identified in (\ref{fin4p}) and give:
\beqn{t0check}
&S_0^* S_2 \sim \overrightarrow{D}^2 + [st + us\log\left(-{u\over s}\right)]( \overrightarrow{\de}_{s}  -  \overrightarrow{\de}_{t})^2 + [su + ts\log\left(-{t\over s}\right)] ( \overrightarrow{\de}_{s}  -  \overrightarrow{\de}_{u})^2 \nonumber \\
&\to (s^2 \overrightarrow{\de}_{s}^2 + u^2 \overrightarrow{\de}_{u}^2 + 2 s u \overrightarrow{\de}_{s}  \overrightarrow{\de}_{u}) + su ( \overrightarrow{\de}_{s}  -  \overrightarrow{\de}_{u})^2 + {{\cal O}}(t) = {{\cal O}}(t) \, ,
\eeqn
where one should stress that the ordering of the derivatives is important for the cancellation. 
\item Extra terms can originate from a careful evaluation of the leading contribution at sub-leading level. An example is the constant $-4$ appearing in (\ref{sumij1}). Also, as already mentioned,  the spectrum $dE^{GW}/d \omega$  depends on which variables are held fixed while the graviton's angular variables are integrated over.
Two of them are obviously the total center-of-mass energy and the graviton's frequency (or, equivalently, $(p_1p_2)$ and $(p_3p_4)$). The angle-integrated spectrum, however, must depend on a third variable somewhat related to the scattering angle of the underlying 4-point function. One symmetric choice would be to keep $(p_1p_3)+(p_2p_4)-(p_1p_4) - (p_2p_3)$ fixed,  but it's by no means unique. In the next section we will see an explicit example in which these extra sub-leading  terms are crucial for carrying out an important check. 
\end{itemize}

We end up this Section by applying the general result (\ref{fin4p})  to two examples of gravitational bremsstrahlung from two-particle collisions. As we will see in the first example the vanishing of the $S_0^* S_2$ term at $t=0$ does not occur. This is because contributions of the type $t^2  \overrightarrow{\de}_{t}^2$ are very sensitive to the amplitude they are acting on. They do not vanish, for instance, if the amplitude has a pole at $t=0$ as in the example presented below. In fact, already the vanishing of $S_2$ in Eq. (\ref{softoper})  for $p_2+p_3  \to 0$ crucially depends on which amplitude the $J_i$ operators act on. In any case there are (tree level) amplitudes for which the above check must hold and they represent a very good test of our final result.

\subsection{Tree-level gravitational scattering}

Let us first consider the tree-level gravitational elastic scattering of two different scalars. In this case, the amplitude receives only a $t$ channel contribution giving (up to an irrelevant numerical factor)
\be{A4}
{\cal M} = - \frac{su}{t}  = \frac{s^2}{t} + s =  \frac{u^2}{t} + u = \frac{su}{s+u}  \, . 
\ee
We will then apply Eq.(\ref{fin4p}) to Eq.(\ref{A4}) by defining: 
\be{Btwo}
 \langle B_2 \rangle_{\cal M}  \equiv \frac{\pi}{2 G \hbar \omega^2} \frac{{\cal M}^{\dagger}B_{2}{\cal M} }{|{\cal M} |^{2}}\, , 
\ee
where, as already explained, we define the derivatives appearing in  $B_{2}$ to act on ${\cal M}$ or ${\cal M}^{\dagger}$ first, {\it i.e.} before any possible multiplier. 
We also recall that the differential operators appearing in (\ref{fin4p}) can be taken to act on any one of the expressions in (\ref{A4}) according to convenience.

We first note that the operators $\overleftarrow{D}^2$ and $\overrightarrow{D}^2$ appearing in (\ref{fin4p}) can be replaced, after the above mentioned ordering of the derivatives, by $\overleftarrow{D} ( \overleftarrow{D}  -1)$ and $\overrightarrow{D}(1- \overrightarrow{D})$, respectively. But since $D$ gives just the $s,t,u$ dimensionality of ${\cal M}$, which is 1, these operators give simply a vanishing result.

Let us now turn to the other two differential operators appearing in Eq.(\ref{fin4p}).
For the operator $\overleftrightarrow{\Delta}^2_{su}$ it is most convenient to use the expression for ${\cal M} $ that does not contain $u$ so that we only need to consider the derivatives w.r.t. $s$. These are readily computed to give, after some simple algebra:
\be{Desu}
{\cal M} ^*\overleftrightarrow{\Delta}^2_{su}{\cal M}  = - \left( \frac{4s^2}{t^2} + 4\frac{s}{t} + 2 \right)  \Rightarrow \frac{{\cal M} ^*\overleftrightarrow{\Delta}^2_{su}{\cal M} }{{\cal M}^*{\cal M} } = - 2\left(\frac{1}{u^2} + \frac{1}{s^2}  \right)\, .
\ee

Similarly, for the operator 
  $\overleftrightarrow{\Delta}^2_{st}$ it is more convenient to work with the $s$-independent form of $A$. This gives:
  \be{Dest}
  {\cal M}^*\overleftrightarrow{\Delta}^2_{st}{\cal M} = 2 \frac{u^3}{t^3} \left(\frac{u}{t} +2\right) \Rightarrow \frac{{\cal M}^*\overleftrightarrow{\Delta}^2_{st}{\cal M}}{{\cal M}^*{\cal M}} =
   2\left(\frac{1}{t^2} - \frac{1}{s^2}  \right)\, .
  \ee

Inserting now (\ref{Desu}) and (\ref{Dest}) in  (\ref{fin4p}) we find the rather elegant result:
\be{B2tree}
 \langle B_2 \rangle_{\cal M} = 2 \left(\frac{1}{t^2} - \frac{1}{s^2}  \right) \left[st + us\log\left(-{u\over s}\right)\right]  - 2 \left(\frac{1}{u^2} + \frac{1}{s^2}  \right)
 \left[su + ts\log\left(-{t\over s}\right)\right] \, .
 \ee
 
 This can then be converted into the following result for the sub-sub-leading correction to the (unpolarized) flux:
 \be{fluxtree}
\frac{d E_2^{GW}}{d \omega} = 4\frac{G (\hbar \omega)^2}{\pi}{{f}}(s, t, u)\, ,
\ee
 where
\be{AdaaA}
{{f}}(s, t, u) =  1+ \frac{s}{t}-\frac{s}{u}+\frac{u}{t} \left( \frac{s}{t} -\frac{t}{s} \right)\log\left(-{u\over s}\right) -\frac{t}{u}  \left( \frac{s}{u} +\frac{u}{s} \right) \log\left(-{t\over s}\right) \, ,
\ee
and, in spite of appearance, there are no singularities at either $t=0$ or $u=0$.
Re-expressing the Mandelstam variables
in terms of 
the cosine of the scattering angle $x \equiv \cos \theta_{s}$,   
\be{stu}
\frac{t}{s} = {1\over 2}(x-1), \,\,\,\frac{u}{s} =  -{1\over 2} (x+1)\, , 
\ee
we can rewrite Eq.(\ref{AdaaA}) as  
$${{f}}(s, t, u)  ={{f}}(x)= 1- \frac{2}{1-x} + \frac{2}{1+ x} - \frac{1+x}{1-x}\left(\frac{2}{1-x} - \frac{1-x}{2} \right) \log\Big(\frac{1+x}{2} \Big)$$
\be{Fx}
+ \frac{1-x}{1+x}\left( \frac{2}{1+x} + \frac{1+x}{2} \right)\log\Big(\frac{1-x}{2} \Big)\, . 
\ee
The function ${{f}}(s, t, u)$ is displayed in Fig.1. It is positive, with a maximum of $3/2$ at $t=0$ (reached with an infinite slope) a value of $1/2$ at $u=0$ and a minimum of about $0.25$ at $x \sim -0.3$. We stress however again that only the sum of the leading and non-leading contributions have an non-ambiguous meaning.
If, for instance, the leading term is defined through eq. (\ref{sumij}), the $O(\omega^2)$ correction is given by 
\begin{equation} \label{AdaaB}
\widehat{f} = {{f}} - 2 = -1+ \frac{s}{t}-\frac{s}{u}+\frac{u}{t} \left( \frac{s}{t} -\frac{t}{s} \right)\log\left(-{u\over s}\right) -\frac{t}{u}  \left( \frac{s}{u} +\frac{u}{s} \right) \log\left(-{t\over s}\right) \, ,
\end{equation} 
because of the $-4$ appearing in eq. (\ref{sumij1}).

\begin{figure}[t]
\centerline{ \includegraphics [height=8cm,width=0.8\columnwidth]{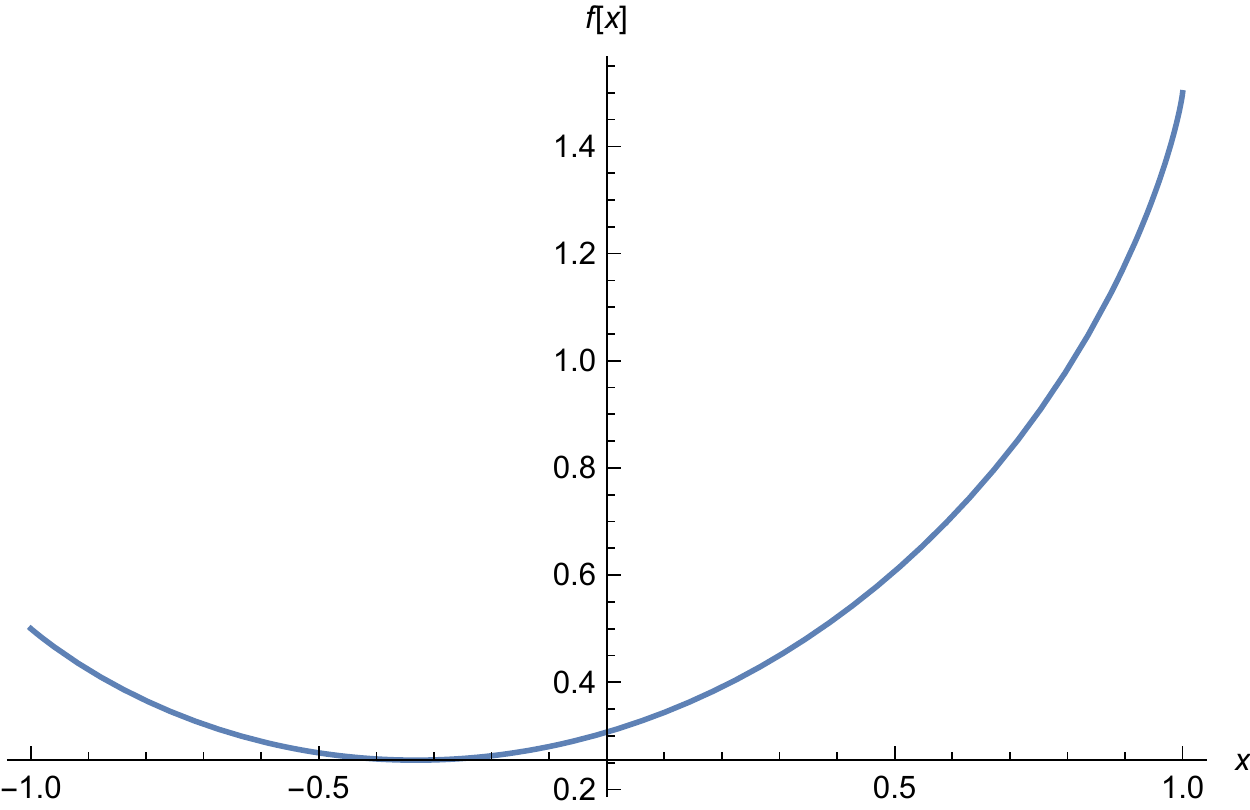}}
\vspace*{-1ex}
\caption{The expression ${{f}}(x)$ appearing in the $B_{2}$ expression is displayed as 
a function of the cosine of the scattering angle $x=\cos \theta_{s}$.  }
\label{plot} 
\end{figure}

As a very non trivial check of our procedure (in particular of our recipe for defining partial derivatives) we present, in Appendix C, an ab-initio tree level calculation of the single-graviton emission amplitude from two-body massless-scalar scattering in ${\cal N} = 8$ supergravity. Quite remarkably, the soft expansion of the result confirms both the vanishing of the ${{\cal O}}(\omega)$ correction to the unpolarized flux and its precise functional form (\ref{AdaaA}) or better (\ref{AdaaB}) at ${{\cal O}}(\omega^2)$.
The full agreement requires the already mentioned careful identification of sub-leading terms implicitly contained in the leading terms of the two calculations. They are the same up to the constant $-4$  appearing in eq. (\ref{sumij1}). The latter is thus an important ingredient for the success of the check.

To summarize, we have found that in this case the leading correction to the ZFL of (\ref{B0Espectrum})  is  of relative order ${(\hbar \omega)^2}{/}{Q^2}$
where $Q \sim \sqrt{-t}$ is the momentum transfer in the process. The correction looks like a quantum effect if we define the classical limit as $\hbar \rightarrow 0$ at fixed $\sqrt{s}$ and $\omega$ (which is the usual way one deals with gravitational wave fluxes in General Relativity). However, if we use the uncertainty principle to replace $Q$ by $\hbar/b$, it becomes a classical-looking correction of relative order $(\omega b)^2$ which is in agreement with expectations about soft-graviton theorems.
 The positive sign may look surprising since, by energy conservation, the constant ZFL value has to leave the way to a decreasing spectrum at higher frequency; however, a maximum at  $\omega \sim 1/b$ has been found to occur \cite{CCV18} in the process discussed in the next subsection.

\subsection{Gravitational scattering in the leading eikonal approximation}

 As a second example we consider the same gravitational scattering in the very high (i.e. transplanckian) energy regime where an all-loop resummation is needed and has been carried out in \cite{Amati:1987uf} (see also \cite{Muzinich:1987in,tHooft:1987vrq}) in the small deflection angle limit. In that case 
the elastic $S$-matrix takes a very simple form in impact-parameter space (we recall that the impact parameter $b$ is related to the orbital angular momentum by
 $b = {2 J}/{\sqrt s}$ and is related by a Fourier transform to the momentum transfer $Q$):
\be{ACVeik}
 {\mathcal S}_{el}(s, b) \approx \exp\left(- i \frac{G s}{\hbar} \log \frac{b^2}{L^2} \right)\, , 
\ee
where $L$ is some infrared cut off screening the (unobservable)  infinite Coulomb phase already well-known in electromagnetic scattering.

Note that this expression has the form of a semiclassical approximation (valid at $\frac{G s}{\hbar} \gg1$) with $Gs$ playing the role of a classical action.
Indeed (\ref{ACVeik}) leads to known classical features of gravitational scattering:
\begin{itemize}
\item The derivative of the exponent w.r.t. $b$  provides the correct gravitational deflection angle (the generalization of Einstein's deflection angle to the case of massless-particle collisions).
\item The derivative of the exponent w.r.t. $E = \sqrt{s}$ gives the (Shapiro) time delay as a function of the impact parameter. Note that while the deflection angle is independent of $L$, the time delay is not. Fortunately what matters are normally time-delay differences for which $L$ again drops out.
\end{itemize}

Let us now apply $B_1$ and $B_2$ to (\ref{ACVeik}) in order to compute the low-energy gravitational radiation accompanying the (otherwise elastic) collision.
The first qualitative remark to be made is that, to leading order in $\hbar$, we can keep only the action of the differential operators on the exponent (just like in the usual WKB approximation). As a result each derivative brings down a $\hbar^{-1}$ factor that precisely compensate for the explicit positive powers of $\hbar$ appearing in ${dE^{GW}}{/}{d\omega}$. In other words we obtain spectra which have a smooth classical limit unlike in the case discussed previously.

At a more quantitative level we find that, also in this case, $B_1$ of eq. (\ref{simpledB1}) gives a vanishing result, when acting on (\ref{ACVeik}). The ``surprise" comes from $B_2$. If we use our final expression (\ref{fin4p})  for $B_2$ and we let the derivatives act just on the exponent we find:
\be{B2onACV}
\frac{\hbar  {\mathcal S}_{el}^{\dagger} B_2  {\mathcal S}_{el}(s, b)}{ {\mathcal S}_{el}^{\dagger}  {\mathcal S}_{el}(s, b)} = \frac{dE_2^{GW}}{d \omega}  = \frac{4G}{\pi} (\hbar \omega)^2 \left( \frac{G s}{\hbar} \log \frac{b^2}{L^2} \right)^2 + \dots = \frac{ Gs}{\pi}  (R_S \omega)^2  \log^2 \frac{b^2}{L^2} + \dots \, , 
\ee
where $R_S = 2 G \sqrt{s}$. 
Thus we recover, as expected, a classical correction to the ZFL. 

Unfortunately, (\ref{B2onACV}) is IR divergent. The divergence comes from the infinite Coulomb phase which is typical of four-dimensional physics.
Such an infinity should disappear from any physical observable meaning that the initial recipe for computing the sub-leading terms has to be modified. 
In recent papers \cite{Laddha:2018rle,Sahoo:2018lxl} the authors have proposed to replace the infrared cutoff $L$ by $\omega^{-1}$.
This recipe has been supported by known classical results \cite{Damourpc} and also, at ${{\cal O}}(\omega)$ level, by the independent IR-singularity free method developed in \cite{CCV18}.

If we apply such a recipe to (\ref{B2onACV}) we get the ${{\cal O}}(\omega^2)$ correction to the (unpolarized) energy flux in the form:
\be{fluxACV}
 \frac{dE_2^{GW}}{d \omega}  = \frac{Gs}{4 \pi} \theta_s^2  ( \omega b)^2  \log^2 (\omega^2 b^2) + \dots \, . 
\ee
This result agrees with the one obtained in \cite{CCV18} up to the overall (positive) constant that cannot be fixed in the small-angle approximation used in \cite{CCV18}. As pointed out there, since (\ref{fluxACV}) represents a positive leading correction to the ZFL, it implies a {\it maximum} of ${dE_2^{GW}}{/}{d \omega}$ displaced from $\omega =0$ by an ${{\cal O}}(1/b)$ amount.

 It also provides another independent check of the validity of the recipe proposed in \cite{Laddha:2018rle,Sahoo:2018lxl}.
 In the next section we will offer some ideas on how one can try to justify that recipe through a suitable modification of the soft theorems themselves.

\section{Elimination of (and finite terms from)  infrared divergences}
\setcounter{equation}{0}

In order to understand the origin of our (hopefully spurious) IR singularity in (\ref{B2onACV}) it is better to step back and consider the sub-leading correction $S_1$ before computing the cross section. Let us also restrict ourselves to $n=4$
 \be{S_1}
 {\cal M}_{5}(p_i; q) \approx \kappa \sum_{i=1}^4  {p_i h J_i q \over qp_i}  {\cal M}_{4}(p_i) \equiv S_1  {\cal M}_{4}(p_i) \, . 
 \ee
 If we use for  ${\cal M}_{4}(p_i)$ the (Fourier transform of) the eikonal expression (\ref{ACVeik}) and let the derivatives present in $J_i$ act on the exponent we find:
  \be{S_1oneik}
 S_1 \approx (-i \kappa) \sum_{i=1}^4  {p_{i \mu} h^{\mu \nu}  q^{\rho} \over qp_i} \sum_{j, \eta_j = \eta_i} \frac{G (p_i p_j - p_j p_i)_{\nu \rho}}{\hbar}  \log \frac{b^2}{L^2} \, , 
 \ee
 where the restriction of the sum over the $i, j$ pairs to be both incoming ($\eta_i = \eta_j = +1$) or outgoing  ($\eta_i = \eta_j = - 1$) comes from the way the Coulomb phase originates (see last section of \cite{Weinberg}). We note that essentially the same term came out from the approach of \cite{Laddha:2017ygw, Laddha:2018rle} where it was argued that the IR cutoff should be replaced by a $\log(\omega^{-1})$.
 
 We see  clearly here the origin of the Coulomb divergence. This propagates to the $S^*_0S_1$ contribution to $B_1$ but cancels with the $S^*_1S_0$ contribution thanks to its over all imaginary phase. We will now argue that such a divergence is spurious and should be cancelled by contributions not included in the naive definition of $S_1$.
 
 Consider indeed two classes of sub-leading contributions according to whether the final soft graviton is emitted from an outgoing or an incoming leg.  In the former case the initial state participating in the $2 \rightarrow 2$ sub-process gives the usual contribution to the Coulomb phase, but the two hard final particles fail to do so since their total momentum is the initial one minus the soft graviton's. The mismatch would produce exactly terms like in (\ref{S_1oneik}) with  $\eta_i = \eta_j = - 1$.
 But, obviously, the mismatch is compensated by the fact that also the final soft graviton contributes to the total Coulomb phase by its own rescattering on the hard final particles.
 For the latter kind of contributions (emission from initial particles) the situation is similar for the final particle's contribution to the phase. For the initial particles contribution one can argue that the Coulomb-divergence comes from the asymptotic (initial and final) states and therefore the initial state to be considered in {\it not} the one entering the blob but the actual initial state carrying the whole energy.
 
 In view of the above it looks that the naive recipe for $S_1$ is wrong on two accounts. The contribution to (\ref{S_1oneik}) with  $\eta_i = \eta_j = +1$ should be simply omitted while the one with  $\eta_i = \eta_j = - 1$ should be supplement with a new term due to the rescattering of the soft graviton on the other final particles.
 We note that a similar contribution also appears in the treatment of \cite{Ciafaloni:2015xsr} and is crucial in order to recover agreement with \cite{Gruzinov:2014moa}.
 
 We may ask whether other modifications of the naive recipe for $S_1$ are needed because of the standard IR divergences. It was argued in \cite{Bern:2014oka} that the correct replacement in that case is:
 \beqn{Bernrepl}
 && (S_0 + S_1)  {\cal M}_{4}^{1-loop-div} = (S_0 + S_1) \sum_{i,j = 1}^{4} \frac{1}{\epsilon} 
 {G s_{ij} \log{ |s_{ij}|\over \mu^2}}  {\cal M}_{4}^{tree} \nonumber \\
  &\rightarrow& S_0  {\cal M}_{5}^{1-loop-div} = S_0 \sum_{i,j = 1}^{5} \frac{1}{\epsilon} {G s_{ij} \log { |s_{ij}|\over \mu^2}} {\cal M}_{5}^{tree}
 \eeqn
 where ${1}{/}{\epsilon}$ plays the role of our $\log L$. This suggests that the true IR divergences are cured as usual by the interplay between real and virtual contributions.
 Furthermore, we will argue that, provided $\Delta E \gg \hbar \omega$ there should be no important finite leftover contribution  to ${dE^{GW}}{/}{d\omega}$.
 
 The argument goes as follows. Consider a generic process in which there is a certain number of hard external particles of characteristic energy $E$ and a number of soft external particles of energy around $\hbar \omega \ll E$, the energy/frequency at which we wish to compute the spectrum of GWs.
 The virtual corrections can be split into those with momenta larger or smaller than $\hbar \omega$. The former contribute only when the gravitons are exchanged between the hard
 particles, the latter contribute to all graviton exchanges. Consider now the real emission corrections to that process distinguishing those with emitted gravitons of energy less than $\hbar \omega$ and those with higher energy up to the $\Delta E$ upper bound. The latter will cancel the IR divergences of the corresponding virtual contributions leaving behind the usual logarithmic dependence on ${E}{/}{\Delta E}$. The crucial point is that, instead, the real gravitons softer than $\hbar \omega$ {\it exactly} cancel their virtual counterpart (since for both the real and the virtual gravitons the upper limit is $\hbar \omega$). 
 Therefore, provided $\Delta E \gg \hbar \omega$, the net result is nothing but the usual infrared factor for the no-emission amplitude which cancels out in ${d E^{GW}}{/}{d\omega}$. Obviously, some dependence on  ${\Delta E}{/}{E}$ will remain as long as that quantity is finite.
 
 In conclusion, while we expect important corrections to the naive recipe to come from the IR divergence related to the Coulomb phase (which has no real counterpart), the usual cancellation mechanism between real and virtual IR divergences should leave room to no substantial correction as long as $\hbar \omega \ll \Delta E \ll E$. A complete and general calculation of the finite corrections coming from the Coulomb phase goes beyond the scope of this paper.

\section{Summary and outlook}
\setcounter{equation}{0}

In this paper we have combined the low-energy theorems for graviton emission, together with some reasonable assumption about how multi soft graviton emission exponentiates to give a coherent state, in order to compute the spectrum of gravitational radiation emitted to leading, sub-leading, and sub-sub-leading order in the frequency.
For simplicity we have only considered the case of massless (equivalently highly relativistic) spin-less particles, and we have summed over the gravitational wave (GW) polarizations and integrated over the angular distribution of the radiation. The results have been expressed therefore in terms of the GW energy flux ${d E^{GW}}{/}{d\omega}$.

At leading order we could easily reproduce the known \cite{Smarr:1977fy} constant zero-frequency limit (ZFL). At sub-leading order we obtained a general result which happens to vanish for GW emission from a two-body collision. For the case of  the unpolarized and/or angle-integrated flux this is in agreement with other results \cite{Laddha:2017ygw,CCV18}.

At sub-sub-leading order (i.e. to ${\cal O}(\omega^2)$) we derive a nice and fairly compact expression for the general case which further simplifies in the case of a two-particle collision process. The expression passes a number of non trivial checks including one against an {\it ab-initio} calculation in ${\cal N} =8$ supergravity.

An important conclusion of our method is that the soft theorems, while presumably valid  at tree-level or in $D>4$, must be amended at higher orders in $D=4$ because of IR problems. 
Actually there are two kinds of IR problems that can invalidate, in principle, the soft-graviton theorems. 

The first is the usual IR catastrophe which is solved by taking into account both real and virtual IR divergences and their cancellation for physically measurable, finite resolution, observables. We have argued, in Section 6, that these divergences should be also harmless as long as  ${d E^{GW}}{/}{d \omega}$ is concerned and the energy-resolution is parametrically larger than $\omega$.

The second kind of divergences comes from the familiar infinite Coulomb phase, also known in QED. Usually, such a phase is unobservable and cancels out when computing physical quantities. However, when the differential (angular-momentum) operators, appearing in the soft theorems at non-leading order, act of such a divergent phase they generate infinities that do not cancel out in ${d E^{GW}}{/}{d \omega}$ \footnote{Possibly, this problem can be cured by  by changing the prescription 
on the S-matrix and asymptotic states as done in QED by Kulish and Feddeev \cite{KF}, 
something beyond the purposes of this paper.}. This IR problem has already been noticed  in recent papers \cite{Sen:2017xjn, Sen:2017nim, Laddha:2017ygw}, whose authors came up with a recipe for dealing with the problem at sub-leading order. Such a recipe has been confirmed by checks against known results and against calculations based on the eikonal method \cite{CCV18} which are free of such problems.

Using the recipe to the sub-sub-leading order we find that the leading correction to the ZFL has exactly the structure found independently in \cite{CCV18} which implies a bump in the GW spectrum at $\omega b \sim 1$. 

For the future it would be clearly important to develop a more rigorous approach to dealing with  IR problems in $D=4$ and,  by reversing the Cachazo-Strominger argument \cite{5}, to understand the outcome in terms of an anomaly in the extended BMS symmetry responsible for the soft-graviton theorems at sub-leading level (see \cite{Strominger:2017zoo} for a recent review).
The other direction of research would be to extend the formalism to massive and/or spinning particles in the initial and final states. Such calculations would nicely complement those presently under way (see e.g. \cite{Bern:2019nnu}) for computing the conservative part of the effective potential at the third post-Minkowskian (3PM) level by providing its dissipative (radiation) counterpart.

\vspace{1cm} 

{\large \bf Acknowledgments}

\vspace{4mm}

We would like to thank  Zvi Bern, Marcello Ciafaloni, Dimitri Colferai, Dario Consoli, Thibault Damour, Paolo Di Vecchia, Maurizio Firrotta, Antonino Marcian\`o, Leonardo Modesto, Matin Mojaza, Massimo Porrati, Ashoke Sen and Nazario Tantalo for useful discussions.
 A.~A. acknowledges support by the National Science Foundation of China (NSFC), through the grant No. 11875113.
  A.~A. also acknowledges Alexey Kavokin and West Lake Institute (Hangzhou, China) for hospitality during the completion of this paper. 
 M.~B. was partially supported by the MIUR-PRIN contract 2015MP2CX4002 {\it ``Non-perturbative aspects of gauge theories and strings''}.

\appendix

\section{Computation of the ``shift terms" arising in the ${{\cal O}}(\omega^{2})$ corrections}
\setcounter{equation}{0}

Let us consider the action of the shift (as defined in Eq.(\ref{shiftdef})) on Eq.(\ref{B2qqW}) dealing separately with the $B_{20}, B_{02}$ and $B_{11}$ factors. 

Before proceeding one has to recall that $\overrightarrow{J}_i$ are differential operators that in principle act on any function of $p_i$ to their right (or left for $\overleftarrow{J}_i$). However, some of the dependence on $p_i$ is `spurious' since it is generated by the shift of the integration variable $q$ into $\tilde{q}_{ij} = q + ...$. In practice the derivatives act only on the explicit dependence on $p_i$ in ${\cal S}_{if}$ or ${\cal S}^*_{if}$ or on the $p_i$ that appear in another $J_i$ as in $S_2$.

\begin{itemize}  
\item
For $B_{11}$ the situation is under control and one has to simply interpret any derivative as only acting on ${\cal S}_{if}$ or ${\cal S}^*_{if}$. Moreover the terms with $i=j$ do not contribute and we find: 
$$(11)=\frac{1}{2}\sum_{i,j}\frac{(p_{i}\overleftarrow{J}_{j}q)(p_{j}\overrightarrow{J}_{i}q)-(p_{i}\overleftarrow{J}_{i}q)(p_{j}\overrightarrow{J}_{j}q)-(p_{i}p_{j})(q\overleftarrow{J}_{i}\overrightarrow{J}_{j}q))}{(qp_{i})(qp_{j})}$$
$$=\frac{1}{2}\sum_{i,j}\frac{(p_{i}\overleftarrow{J}_{j}\tilde{q})(p_{j}\overrightarrow{J}_{i}\tilde{q})-(p_{i}\overleftarrow{J}_{i}\tilde{q})(p_{j}\overrightarrow{J}_{j}\tilde{q})-(p_{i}p_{j})(\tilde{q}\overleftarrow{J}_{i}\overrightarrow{J}_{j}\tilde{q}))}{(qp_{i})(qp_{j})}$$
\be{shifts}
-\frac{3}{2}\sum_{i\neq j}(p_{i}\overleftarrow{\partial}_{i})(p_{j}\overrightarrow{\partial}_{j})+\frac{5}{2}\sum_{i}(p_{i}\overleftarrow{\partial}_{i})(p_{i}\overrightarrow{\partial}_{i})\, , 
\ee
which, by rewriting the $\sum_{i\neq j}$ as a sum all over $i,j$ and subtracting to it the diagonal terms, can be rewritten as
\be{rewritten}
{\rm shift}(11)=-\frac{3}{2}\sum_{i,j}(p_{i}\overleftarrow{\partial}_{i})(p_{j}\overrightarrow{\partial}_{j})+4\sum_{i}(p_{i}\overleftarrow{\partial}_{i})(p_{i}\overrightarrow{\partial}_{i})\, ,
\ee
The $(-3/2)$ factor in Eq.(\ref{shifts}) comes from three $-1/2$ factor for each of the three initial terms 
in $(11)$; the factor $(+5/2)$ is the sum of a coefficient $1$ from the first, a coefficient $0$ from the second
and a coefficient $+3/2$ from the third. 

\item
For $B_{20}$ and $B_{02}$ one has to be careful and take into account that, for $i=j$, 
$$J^i_{\mu\nu} J^i_{\rho\sigma} =  p^i_\mu p^i_\rho \de^i_\nu \de^i_\sigma + \eta_{\nu\rho} p_\mu\de_\sigma - [\mu\nu] - [\rho\sigma] + ([\mu\nu], [\rho\sigma])$$ \, .
We find: 
$$
(20)+(02)=-\frac{1}{2}\Big[\sum_{i,j}\frac{(q\overleftarrow{J}_{j}p_{i})(p_i\overleftarrow{J}_{j}q)}{(qp_{i})(qp_{j})}+(i\leftrightarrow j ) \& (\leftrightarrow)  \Big]
$$
\be{lala}
=-\frac{1}{2}\Big[ \sum_{i\neq j}\frac{(\tilde{q}\overleftarrow{J}_{j}p_{i})(p_{i}\overleftarrow{J}_{j}\tilde{q})}{(qp_{i})(qp_{j})}+(i\leftrightarrow j)  \& (\leftrightarrow) \Big] +2\sum_{i}\Big[(p_{i}\overleftarrow{\partial}_{i})^{2}+(p_{i}\overrightarrow{\partial}_{i})^{2}\Big]\, . 
\ee
where $(\leftrightarrow)$ means reversing the arrow-orientation. 
The last term of Eq.(\ref{lala}) is the sum of  two shift terms: from 
$i=j$  (giving a numerical factor $1/2$) and from $i\neq j$ (providing a $3/2$ factor). 

\end{itemize}  

Finally, combining $(02)+(20)+(11)$, 
we find: 
\be{hjak}
{\rm Total\, Shift}=-\frac{3}{2}\sum_{i,j}(p\overleftarrow{\partial}_{i})(p_{j}\overrightarrow{\partial}_{j})+2\sum_{i}(p_{i}\overleftarrow{\partial}_{i}+p_{i}\overrightarrow{\partial}_{i})^{2}\, ,
\ee
where the last term is obtained by combining the last term of Eq.(\ref{rewritten}) with the last term of Eq.(\ref{lala}). This is the result we inserted in
(\ref{B2new}) modulo the fact that, according to our ordering of the derivatives, the second term in (\ref{hjak}) has to be ordered accordingly.

\section{Non covariant calculation of $P_{\mu} P_{\nu} \tilde{M}^{\mu \nu}$ }
\setcounter{equation}{0}

 Relying on Lorentz invariance, one can compute the relevant integral in the CM frame whereby
\beqn{PMP}
P_{\mu}P_{\nu} \widetilde{M}_{ij}^{\mu\nu}  
&=&\int {d^3q\over |q| qp_i qp_j} \delta\Big({qP\over \Lambda^2} + 1\Big) {\left(qP - {qp_i p_jP\over p_i p_j}
- {qp_j p_iP\over p_i p_j}\right)}^2 \nonumber \\
&=& {\Lambda^4 \over s} {s \over E_i E_j (1 -v_i v_j)^2} \int d\Omega
{ [n(v_i+v_j) -1 -v_iv_j]^2 \over (1-nv_i) (1-nv_j)} \, .
\eeqn

The angular integral can be rewritten as
\be{expr}
\int d\Omega
{ [n(v_i+v_j) -1 -v_iv_j]^2 \over (1-nv_i) (1-nv_j)} = (1+v_iv_j)^2 M_{ij}^{0}- 2(1+v_iv_j) M_{ij}^{a}(v^a_{i} + v^a_{j}) +M_{ij}^{ab} (v^a_{i} + v^a_{j}) (v^b_{i} + v^b_{j})\, .
\ee
The first two integrals were already computed, for $B_{0,1}$  in the `covariant' approach in Section 3. 
In the mass-less limit under consideration, after introducing a small mass regulator, one finds
$$
M_{ij}^{0} = {4\pi \over (1-v_iv_j)} \log \left({1-v_iv_j \over 2} {2|E_i| \over m_i}{2|E_j| \over m_j}\right)
$$
and 
$$
M_{ij}^{a} = {1 \over 1-(v_iv_j)^2} \left\{ v_i^a[M_{ij}^{0} (1-v_iv_j) + v_iv_j L_i - L_j] + (i{\leftrightarrow}j) \right\}\, , 
$$
where (in the mass-less limit) 
$$
L_i = 4\pi \log {2|E_i| \over m_i}\, , 
$$
so that 
$$
M_{ij}^{a}(v^a_{i} + v^a_{j}) = 2 M_{ij}^{0} - L_i - L_j\, .
$$

Thanks to the shift, as noticed in Section 3, the Lorentz invariant final result is free of divergences in the massless limit $m_i \to 0$. This requires cancellations between $L_i,L_j$ and the log terms in $M_{ij}^{0}$.

The $M^{ab}_{ij}$ integral matrix can be decomposed as
\be{Iabn}
M_{ij}^{ab}=\int d^{2} n \frac{n^{a}n^{b}}{(1-n\cdot v_{i})(1-n\cdot v_{j})}=\alpha \delta^{ab}+\beta(v_{i}^{a}v_{i}^{b}+v_{j}^{a}v_{j}^{b})
+\gamma(v_{i}^{a}v_{j}^{b}+v_{j}^{a}v_{i}^{b})\, .
\ee
In order to determine the `scalar' integrals $\alpha, \beta, \gamma$, one can project $M^{ab}_{ij}$ along the three independent components:
\be{deltaab}
[1]: \qquad \delta_{ab}M_{ij}^{ab}=M_{ij}^{0}= I =3\alpha+\beta(v_{i}^{2}+v_{j}^{2})+2\gamma v_{i}\cdot v_{j}\, , 
\ee
\be{vivj}
[2]: \qquad (v_{i}^{a}v_{j}^{b}+v_{j}^{a}v_{i}^{b})M^{ab}_{ij}=2\int d^{2}n \frac{nv_{i}nv_{j}}{(1-nv_{i})(1-nv_{j})}
\ee
$$=2(4\pi)-2\int \frac{d^{2}n}{(1-nv_{i})}-2\int \frac{d^{2}n}{(1-nv_{j})}+2\int \frac{d^{2}n}{(1-nv_{i})(1-nv_{j})}$$
$$=8\pi-2(L_{i}+L_{j})+2 M_{ij}^{0} = 2[4\pi-(L_{i}+L_{j})+M_{ij}^{0}] = 2 J $$ $$=2v_{i}v_{j}\left[\alpha+\beta(v_{i}^{2}+v_{j}^{2})+\gamma\left(v_{i}v_{j}+\frac{v_{i}^{2}v_{j}^{2}}{v_{i}v_{j}} \right)\right]\, ,$$
\be{vivi}
[3]: \qquad (v_{i}^{a}v_{i}^{b}+v_{j}^{a}v_{j}^{b})M^{ab}_{ij}=\int d^{2}n \frac{(nv_{i})^{2}+(nv_{j})^{2}}{(1-nv_{i})(1-nv_{j})}
\ee
$$=\int d^{2}n \left[ \frac{1-nv_{i}}{1-nv_{j}}+\frac{1-nv_{j}}{1-nv_{i}} - 2 \frac{1}{1-nv_{i}}-2\frac{1}{1-nv_{j}} +\frac{2}{(1-nv_{i})(1-n v_{j})}\right]$$ $$=2M_{ij}^{0}-2(L_{i}+L_{j})+X_{i/j}+X_{j/i} = _{|v_i|=|v_j|=1} 2 K = 
2M_{ij}^{0}-(1+v_iv_j) (L_{i}+L_{j}) + 8\pi (1+v_iv_j)$$
$$=\alpha (v_{i}^{2}+v_{j}^{2})+\beta [(v_{i}^{2})^{2}+2(v_{i}v_{j})^{2}+(v_{j}^{2})^{2}]
+2\gamma(v_{i}^{2}v_{i}v_{j}+v_{j}^{2}v_{i}v_{j})\, ,$$
where, choosing reference frame so that $v_j=|v_j|(0,0,1)$, $v_i=|v_i|(\sin\theta_{i,j}, 0, \cos\theta_{i,j})$ and 
$n=(\sin\theta_{n,j}\cos\phi_{n,j}, \sin\theta_{n,j}\sin\phi_{n,j},\cos\theta_{n,j})$,
$$X_{i/j}=\int d^{2}n \frac{1-nv_{i}}{1-nv_{j}} =\int d\cos \theta_{n,j} d\phi_{n,j}
\frac{1-|v_{i}|\cos \theta_{n,i}}{1-|v_{j}|\cos \theta_{n,j}}$$
$$=\int d\cos \theta_{nj}d\phi_{nj} \frac{1-|v_i|(\cos\theta_{n,j}\cos\theta_{i,j} + [\sin\theta_{n,j}\cos\phi_{n,j}\sin\theta_{i,j}])}{1-|v_{j}|\cos \theta_{n,j}}= 4\pi {v_iv_j\over |v_j|^2} + L_j \left(1 - {v_iv_j\over |v_j|^2}\right)\, .$$

In order to compute Integral matrix, we need to calculate 
\be{Mvv}
M_{ij}^{ab} (v^a_{i} + v^a_{j}) (v^b_{i} + v^b_{j}) = 2(1+c) [\alpha + (1+c)(\beta +\gamma)]\, .
\ee
After some long and tedious algebra one gets
\be{Mvv2}
M_{ij}^{ab} (v^a_{i} + v^a_{j}) (v^b_{i} + v^b_{j}) = 4 M_{ij}^{0} -  (3+c) (L_i+L_j) + 8\pi (1+c)\, .
\ee

Combining with the other two terms one finally gets
\be{PMP}
P_{\mu}P_{\nu} \widetilde{M}_{ij}^{\mu\nu}  = {\Lambda^4 E_i E_j \over (p_i p_j)^2}\left\{ 
(1-v_iv_j)^2 M_{ij}^{0} - (1-v_iv_j)  (L_i +L_j) +8\pi(1+v_iv_j)]\right\} \, .
\ee

As a check notice that the result vanishes, as expected, for $v_i=-v_j$ {\it i.e.} $v_iv_j=-1$ since in this limit $M_{ij}^{0}\to L_i +L_j$. Moreover for $v_i=v_j$ {\it i.e.} $v_iv_j=1$, as expected, one gets a constant $16\pi$.
Using the explicit expressions for $M_{ij}^{0}$, $L_i$, $L_j$ one can check Lorentz invariance and finiteness (as $m_i \to 0$) for generic values of $v_iv_j$. Indeed the result can be written as 
\beqn{PMPfin1}
&&P_{\mu}P_{\nu} \widetilde{M}_{ij}^{\mu\nu} = {\Lambda^4 E_i E_j \over (p_i p_j)^2}
\left\{ 
2\pi (1-v_iv_j) \log\left({1-v_iv_j\over 2}\right)^2+8\pi(1+v_iv_j)]\right\} \nonumber \\
&=& {8\pi  \omega_0^2 \over p_i p_j}
\left\{ {P^2\over 2} \log {P^2 p_i p_j \over 2 Pp_i Pp_j} + \tilde{s}_{ij} \right\}\, .
\eeqn

\section{A non-trivial check in ${\cal N} =8$ supergravity}

\setcounter{equation}{0}
In this appendix, we study the soft limit of the 4-scalar + 1 graviton amplitude
in order to compare the exact results with the ones obtain from soft theorems. 
We will show the perfect agreement between the two approaches provided one takes carefully into account all sub-leading terms, including those somehow hidden below the leading one.

In addition to the fermions, the ${\cal N}=8$ SUGRA multiplet contains the graviton $h_{\mu\nu}$, which is a singlet of the $SU(8)$ R-symmetry, 28 gravi-photons $A_\mu^{[IJ]}$ and 70 massless scalars $\phi^{[IJKL]}$, with $I,J,...=1,...8$. The simplest 4-scalar-1-graviton amplitude involves two pairs of complex conjugate scalars: $\phi = \phi^{1234} $, ${\chi}= \phi^{3456}$, $\bar\phi = \bar\phi_{1234}= \phi^{5678}$ and $\bar{\chi}= \bar{\chi}_{3456}= \phi^{1278}$. 

Relying on the helicity spinor formalism and on the remarkable properties of MHV and SUSY related amplitudes (see {\it e.g.} \cite{Bianchi:2008pu, Elvang:2013cua}),  one finds ($8\pi G = 1$) 
\begin{equation}
{\cal M}_4(\phi(1),{\chi}(2),\bar{\chi}(3),\bar\phi(4)) = {{\cal M}^{MHV}_4(1^-2^-3^+4^+) \over {\langle}12{\rangle}^8}  {\langle}12{\rangle}^2 {\langle}13{\rangle}^2 {\langle}24{\rangle}^2 {\langle}34{\rangle}^2  \nonumber
\end{equation}
\begin{equation}
= {s_{12}  \over {\langle}12{\rangle}^8} {{\langle}12{\rangle}^4 {\langle}12{\rangle}^2 {\langle}13{\rangle}^2\over {\langle}12{\rangle}{\langle}24{\rangle}{\langle}34{\rangle}{\langle}31{\rangle}} {{\langle}12{\rangle}^4 {\langle}24{\rangle}^2 {\langle}34{\rangle}^2\over {\langle}21{\rangle}{\langle}14{\rangle}{\langle}34{\rangle}{\langle}32{\rangle}} = s_{12} {{\langle}13{\rangle} {\langle}24{\rangle} \over {\langle}14{\rangle} {\langle}32{\rangle}}\, ,
\end{equation}
with ${\cal M}^{MHV}_4(1^-2^-3^+4^+)$ the well-known MHV amplitude of $\mathcal{N}=4$ SYM theory. 
Multiplying numerator and denominator by $[14]$ and using momentum conservation ($|4{\rangle}[4| = - |1{\rangle}[1| - |2{\rangle}[2| - |3{\rangle}[3|$) 
so that ${\langle}24{\rangle}[14] = - {\langle}23{\rangle}[13]$, we find 
\begin{equation}
{\cal M}_4(\phi(1),{\chi}(2),\bar{\chi}(3),\bar\phi(4)) = {s_{12}s_{24}\over s_{14}} = {s u \over t} 
 \end{equation}
 which is exactly the same result as for purely gravitational scattering with $s=s_{12}$, $t=s_{14}$ and $u=s_{24}$, thanks to the ordering of the external scalar legs (i.e. 4=1$^*$ and 3=2$^*$).

 The 4-scalar+1-graviton MHV-like amplitude is given by
\begin{equation}
{\cal M}_5(\phi(1),{\chi}(2),\bar{\chi}(3),\bar\phi(4), h^+(5)) = {{\cal M}_5(1^-2^-3^+4^+5^+)\over {\langle}12{\rangle}^8}{\langle}12{\rangle}^2{\langle}24{\rangle}^2{\langle}34{\rangle}^2{\langle}13{\rangle}^2  \, , 
\end{equation}
 where ${\cal M}$ is the 5-graviton MHV amplitude, which can be rewritten, using KLT relation, in terms of MHV amplitudes ${\cal A}_5^{L}$, ${\cal A}_5^{R}$ in $\mathcal{N}=4$ SYM as follows: 
\begin{equation}
{\cal M}_5(1^-2^-3^+4^+5^+) = s_{12}s_{34} {\cal A}^L_5(1^-2^-3^+4^+5^+){\cal A}^R_5(2^-1^-4^+3^+5^+) +  \nonumber
\end{equation}
\begin{equation}
+s_{13}s_{24} {\cal A}^L_5(1^-3^
+2^-4^+5^+){\cal A}^R_5(3^+1^-4^+2^-5^+)=  \nonumber
\end{equation}
\begin{equation}
= s_{12}s_{34} {{\langle}12{\rangle}^4\over {\langle}12{\rangle}{\langle}24{\rangle}{\langle}34{\rangle}{\langle}35{\rangle}{\langle}51{\rangle}} {{\langle}12{\rangle}^4\over {\langle}21{\rangle}{\langle}13{\rangle}{\langle}34{\rangle}{\langle}45{\rangle}{\langle}52{\rangle}}+  \nonumber \end{equation} 
 \begin{equation}
+ s_{13}s_{24} {{\langle}12{\rangle}^4\over {\langle}14{\rangle}{\langle}42{\rangle}{\langle}23{\rangle}{\langle}35{\rangle}{\langle}51{\rangle}}{{\langle}12{\rangle}^4\over {\langle}41{\rangle}{\langle}13{\rangle}{\langle}32{\rangle}{\langle}25{\rangle}{\langle}54{\rangle}}\, . 
\end{equation}

 At five points MHV-like and anti-MHV-like are complex conjugate and different in general. Let us compute 
 ${\cal M}_5(\phi(1),{\chi}(2),\bar{\chi}(3), \bar\phi(4), h^+(5))$. Relying once again of the remarkable properties of MHV and SUSY related amplitudes in the helicity spinor formalism, one finds

\begin{equation}
{\cal M}_5(\phi(1),{\chi}(2),\bar{\chi}(3),\bar\phi(4), h^+(5)) =   \nonumber 
\end{equation}
\begin{equation} = {[15][25][35][45] \over s_{15} s_{25}s_{35}s_{45}}  {{\langle}12{\rangle}{\langle}24{\rangle}{\langle}43{\rangle}{\langle}31{\rangle}\over {\langle}14{\rangle}{\langle}23{\rangle}} 
\{ {\langle}14{\rangle}{\langle}23{\rangle} [12][43] - {\langle}12{\rangle}{\langle}43{\rangle} [13][24] \}\, .
\end{equation}

The differential cross-section obtains 
after mod-squaring the amplitudes and summing over helicities.
The sum over graviton polarisations (still keeping $8\pi G = 1$) yields 
\begin{equation}
\sum_{s=\pm 2} |{\cal M}_5(\phi,\chi,\bar\chi,\bar\phi, h^s)|^2  =  32
{\prod_{i=1}^4 s_{i5} \over (\prod_{i=1}^4 s_{i5})^2} {s_{12}s_{43}s_{13}s_{24} \over s_{14}s_{23}} \varepsilon(1,2,3,4)^2
\end{equation}
after using $|i\rangle[i|=p_i$ to recast spinor products into scalar products.
Now using
\begin{equation} 
\varepsilon^{\mu\nu\rho\sigma} \varepsilon_{\mu'\nu'\rho'\sigma'} = - \delta^{[\mu}_{[\mu'}
\delta^{\nu}_{\nu'}\delta^{\rho}_{\rho'}\delta^{\sigma]}_{\sigma']}
\end{equation}
yields 
\begin{equation}
\varepsilon(1,2,3,4)^2 = {\cal F}( x= s_{12}s_{43}, y= s_{14}s_{23}, z= s_{13}s_{24})  \nonumber
\end{equation}
\begin{equation} =- s_{12}^2s_{34}^2 - s_{13}^2 s_{24}^2 - s_{14}^2s_{23}^2 
+ 2 s_{12}s_{43}s_{13}s_{24} + 2 s_{13}s_{24}s_{14}s_{23} + 2 s_{14}s_{23}s_{12}s_{43}\, 
\end{equation}
known also as the `fake square' (ubiquitous in 3-body phase-space and in 4-pt correlation functions in CFT).
And finally 
\begin{equation}
\sum_{\pm} |{\cal M}_5(\phi_1,\chi_2,\bar\chi_3,\bar\phi_4, h^\pm_5)|^2  = 
-  2 {s_{12}s_{34}s_{13}s_{24} \over s_{14}s_{23} \prod_{i=1}^4 s_{i5}}\times   \nonumber
 \end{equation}
 \begin{equation}
\{s_{12}^2s_{34}^2 + s_{13}^2 s_{24}^2 + s_{14}^2s_{23}^2 
- 2 s_{12}s_{43}s_{13}s_{24} - 2 s_{13}s_{24}s_{14}s_{23} - 2 s_{14}s_{23}s_{12}s_{43}\}\, . 
\end{equation}

In order to proceed it is convenient to set 
\begin{equation}
s_{12} + s_{34} = 2s  \quad , \quad  s_{14} + s_{23} = 2t \quad , \quad s_{24} + s_{13} = 2u   \, , 
\end{equation}
in such a way that $s+t+u = 0 = q^2$, even before the soft limit\footnote{Note that these are precisely the variables $-\Delta_s^2,-\Delta_t^2,-\Delta_u^2$ introduced in (\ref{defDeltas}). For simplicity of notation we denote them here by $s,t,u$ although they are not to be confused with the same variables in the 4-point-function.} . 

One can then express the scalar products in terms of $s,t,u$ and $Q_i = qp_i/2$
\begin{equation}
s_{12} = s + Q_1 + Q_2 - Q_3 - Q_4  \qquad 
s_{34} = s - Q_1 - Q_2 + Q_3 + Q_4 \, , 
\end{equation}
\begin{equation}
s_{14} = t + Q_1 + Q_4 - Q_2 - Q_3  \qquad 
s_{23} = t - Q_1 - Q_4 + Q_2 + Q_3 \, , 
\end{equation}
\begin{equation}
s_{13} = u + Q_1 + Q_3 - Q_2 - Q_4  \qquad 
s_{24} = u - Q_1 - Q_3 + Q_2 + Q_4 \, . 
\end{equation}
in such a way that 
\begin{equation}
s_{12}s_{34} = s^2 - (Q_1 + Q_2 - Q_3 - Q_4)^2  \equiv  s^2 - Q_{12,34}^2
\end{equation}
\begin{equation}
s_{13}s_{24} = t^2 - (Q_1 - Q_3 + Q_2 + Q_4)^2 \equiv  t^2 - Q_{13,24}^2 
\end{equation}
\begin{equation}
s_{14}s_{23} = u^2 - (Q_1 - Q_4 + Q_2 + Q_3)^2 \equiv  u^2 - Q_{14,23}^2 \, . 
\end{equation}

Finally, observing that ${\cal F}(s^2,t^2,u^2) = 0$ since $s+t+u=0$, the non-polarised square-amplitude reads
$$
|{\cal M}|_{n-p}^2 = -  {1\over 2^7} {(s^2 - Q_{12,34}^2)(u^2 - Q_{13,24}^2) \over  Q_1Q_2Q_3Q_4 (t^2 - Q_{14,23}^2)}\times
$$
$$
\{ 2s^2 (  Q_{14,23}^2 + Q_{13,24}^2 - Q_{12,34}^2) +2 t^2 (  Q_{12,34}^2 + Q_{13,24}^2 - Q_{14,23}^2) 
+  2 u^2 (  Q_{14,23}^2 + Q_{12,34}^2 - Q_{13,24}^2)
$$
\begin{equation}+ Q_{12,34}^4 + Q_{13,24}^4 + Q_{14,23}^4 - 
2 Q_{12,34}^2  Q_{13,24}^2 -2 Q_{14,23}^2 Q_{12,34}^2 - 2Q_{14,23}^2 Q_{13,24}^2\}\, . 
\end{equation}
In fact the second line can be further simplified since 

\begin{equation}
Q_{13,24}^2 + Q_{14,23}^2 - Q_{12,34}^2 = -8[Q_1Q_2 + Q_3Q_4]= -2(qp_1qp_2 + qp_3qp_4)
\end{equation} 
\begin{equation}
Q_{12,34}^2 + Q_{14,23}^2 - Q_{13,24}^2 = -8[Q_1Q_3 + Q_2Q_4] = -2(qp_1qp_3 + qp_2qp_4)
\end{equation}
\begin{equation}
Q_{13,24}^2 + Q_{12,34}^2 - Q_{14,23}^2 = -8[Q_1Q_4 + Q_2Q_3] = -2(qp_1qp_4 + qp_2qp_3)\, . 
\end{equation}

To lowest order in $q$ one thus gets \begin{equation}
\label{N8leading}
|{\cal M}|_0^2 =  {1\over 2}{s^2 u^2\over  t^2} {s^2(qp_1qp_2{+}qp_3qp_4){+}t^2(qp_1qp_4{+}qp_2qp_3)
{+}u^2(qp_1qp_3{+}qp_2qp_4)\over qp_1qp_2qp_3qp_4}
\end{equation}
which is precisely what is needed to reproduce Weinberg's ZFL. 

Furthermore, this ``leading" term can be compared with the analogous leading term in
 (\ref{sumij}). 
 It turns out that the following identity holds {\it before} taking the ZFL 
 \beq
{s^2(qp_1qp_2{+}qp_3qp_4) +t^2(qp_1qp_4{+}qp_2qp_3)
+ u^2(qp_1qp_3{+}qp_2qp_4)\over qp_1qp_2qp_3qp_4} =  2  \sum_{i,j}  {(p_i p_j)^2  \over qp_i qp_j } 
\eeq
so that the only difference between (\ref{sumij}) and (\ref{N8leading}) is in the additional $-4$ appearing in the former.

To next `linear' order in $q$ one gets exactly zero, even before integrating over the (soft) on-shell momentum $q$
\begin{equation}
|{\cal M}|_1^2 = 0\, . 
\end{equation}
The next non-trivial contribution is of order $q^2$ and reads
\begin{equation}
|{\cal M}|_2^2 = -  {1\over 8} {s^2 u^2\over  t^2 qp_1qp_2qp_3qp_4} ({{\cal U}} + {\cal V}) 
\end{equation}
where
\begin{equation}
{{\cal U}}= 4 \left[{Q_{12,34}^2\over  s^2}{+}{Q_{13,24}^2\over  u^2}{-}{Q_{14,23}^2\over  t^2}\right][s^2(qp_1qp_2 + qp_3qp_4)+t^2(qp_1qp_4 + qp_2qp_3) +u^2(qp_1qp_3 + qp_2qp_4)] \end{equation}
and
\begin{equation}
{\cal V} = Q_{12,34}^4 + Q_{13,24}^4 + Q_{14,23}^4 - 
2 Q_{12,34}^2  Q_{13,24}^2 -2 Q_{14,23}^2 Q_{12,34}^2 - 2Q_{14,23}^2 + Q_{13,24}^2\, , 
\end{equation}
which is once again a `fake square' ${\cal F}(x=Q_{12,34}^2, y=Q_{14,23}^2, z=Q_{12,34}^2)$ that now, contrary to ${\cal F}(s^2, t^2, u^2)=0$, does not vanish, since there is no choice of signs such that $\pm Q_{12,34}\pm Q_{14,23}\pm Q_{13,24} =0$.

The expression for ${\cal V}$ can be further simplified to
 \begin{equation}
{\cal V} = -4[ (qp_1)^2(qp_2qp_4{+}qp_3qp_4{+}qp_3qp_2) + (qp_2)^2(qp_1qp_4{+}qp_3qp_4{+}qp_3qp_1)  \nonumber
\end{equation}
\begin{equation}
+ (qp_4)^2(qp_1qp_2{+}qp_2qp_3{+}qp_3qp_1) + (qp_3)^2(qp_1qp_4{+}qp_4qp_2{+}qp_2qp_1)]
\end{equation}
so that 
 \begin{equation}
 {{\cal V}\over  qp_1qp_2qp_3qp_4} = -4 \sum_{i\neq j} {qp_i\over qp_j} = - 4 \sum_j {-qp_j\over qp_j} = 16\, . 
 \end{equation}

By the same token, one finds 

  \begin{equation}
 {{{\cal U}}\over  qp_1qp_2qp_3qp_4}  = - 4 \left[C_1 + {s^2\over t^2} C_{s/t} + {u^2\over t^2} C_{u/t} + 
 {s^2\over u^2} C_{s/u} + {u^2\over s^2} C_{u/s}+ {t^2\over s^2} C_{t/s} + {t^2\over u^2} C_{t/u}\right]\, , 
  \end{equation}
where (recall $Q_i = qp_i/2$)
  \begin{equation}
C_1= 2 \left[ \left({Q_1\over Q_4} + {Q_4\over Q_1}\right)+ \left({Q_2\over Q_3} + {Q_3\over Q_2}\right)\right] \, , 
\end{equation}
 \begin{equation}
 C_{s/t} = - 2 - \left({Q_1\over Q_4} + {Q_4\over Q_1}\right)- \left({Q_2\over Q_3} + {Q_3\over Q_2}\right) - 
\left({Q_1Q_2\over Q_3Q_4} + {Q_3Q_4\over Q_1Q_2}\right) \, , 
 \end{equation}
 \begin{equation}
 C_{u/t} = - 2 - \left({Q_1\over Q_4} + {Q_4\over Q_1}\right)- \left({Q_2\over Q_3} + {Q_3\over Q_2}\right) - 
\left({Q_1Q_3\over Q_2Q_4} + {Q_2Q_4\over Q_1Q_3}\right)\, , 
 \end{equation}
 \begin{equation}
 C_{s/u} = 2 + \left({Q_1\over Q_3} + {Q_3\over Q_1}\right)+ \left({Q_2\over Q_4} + {Q_4\over Q_2}\right) + 
\left({Q_1Q_2\over Q_3Q_4} + {Q_3Q_4\over Q_1Q_2}\right)\, , 
\end{equation}
 \begin{equation}
 C_{u/s} = 2 + \left({Q_1\over Q_2} + {Q_2\over Q_1}\right)+ \left({Q_3\over Q_4} + {Q_4\over Q_3}\right) + 
\left({Q_1Q_3\over Q_2Q_4} + {Q_2Q_4\over Q_1Q_3}\right)\, , 
 \end{equation}
 \begin{equation}
 C_{t/s} = 2 + \left({Q_1\over Q_2} + {Q_2\over Q_1}\right)+ \left({Q_4\over Q_3} + {Q_3\over Q_4}\right) + 
\left({Q_2Q_3\over Q_1Q_4} + {Q_1Q_4\over Q_2Q_3}\right) \, , 
 \end{equation}
 \begin{equation}
 C_{t/u}  = 2 + \left({Q_1\over Q_3} + {Q_3\over Q_1}\right)+ \left({Q_2\over Q_4} + {Q_4\over Q_2}\right) + 
\left({Q_1Q_4\over Q_2Q_3} + {Q_2Q_3\over Q_1Q_4}\right) \, , 
 \end{equation}

We would like to determine the soft graviton spectrum at this order ($\omega^2$). To this end one should integrate expressions of the form 
\begin{equation}
\int {d^3q\over |q|} \left\{ C_0 + C_{a/b} {Q_a\over Q_b} + C_{ab/cd} {Q_a Q_b \over Q_c Q_d} \right\}\, . 
\end{equation}

Integrating over the graviton phase space, from the ${\mathcal{U}}$ term, one gets:
\begin{equation}
\int {d^3q \over |q|} {{{\cal U}}\over \prod_i qp_i} = - 4 \left[ - 8(4\pi\omega^2) + \left({s^2\over u^2} - {s^2\over t^2}\right)(p_1\widetilde{M}_{34} p_2 + p_4\widetilde{M}_{12} p_3) \right. \quad  \nonumber
\end{equation}
\begin{equation}
\left. +\left({u^2\over s^2} - {u^2\over t^2}\right) (p_1\widetilde{M}_{24} p_3 + p_4\widetilde{M}_{13} p_2)
        +\left({t^2\over s^2} + {t^2\over u^2}\right) (p_1\widetilde{M}_{23} p_4 + p_2\widetilde{M}_{14} p_3) \right]
\end{equation}
where
\begin{equation}
\widetilde{M}_{ij}^{\mu\nu}= \int {d^3q\over |q|} {\tilde{q}_{ij}^\mu \tilde{q}_{ij}^\nu \over qp_i qp_j} \delta	\Big({qP\over \Lambda^2} - 1\Big) 
\end{equation}
and $4\pi\omega^2$ in the very first term comes from 
\begin{equation}
\int {d^3q\over |q|} \delta	\Big({qP\over \Lambda^2} - 1\Big) =  4\pi\omega^2\, . 
\end{equation}

The results for the various relevant integrals read 
\begin{equation}
p_1\widetilde{M}_{34} p_2 = p_4\widetilde{M}_{12} p_3  = {8\pi \omega^2\over s} \times  \left(-{u t\over s}\right)\, , 
\end{equation}
\begin{equation}
p_1\widetilde{M}_{24} p_3 = p_4\widetilde{M}_{13} p_2  = 
{16\pi \omega^2 \over s t} \left[ 1 + {u\over t} \log\left(-{u\over s}\right)\right] 
\left(-{s t\over u}\right)\left({s t\over u}\right) - {8\pi \omega^2 \over t} \log\left(-{u\over s}\right)\left(-{s t\over u}\right)\, ,  \nonumber
\end{equation} 
\begin{equation}
p_1\widetilde{M}_{23} p_4 = p_3\widetilde{M}_{14} p_2  = 
{16\pi \omega^2 \over s u} \left[ 1 + {t\over u} \log\left(-{t\over s}\right)\right] 
\left(-{s u\over t}\right)\left({s u\over t}\right) - {8\pi \omega^2 \over u} \log\left(-{t\over s}\right)\left(-{s u\over t}\right)\, ,  \nonumber
\end{equation} 
that can be written as 
\begin{equation}
p_1\widetilde{M}_{34} p_2 = p_4\widetilde{M}_{12} p_3  = -{8\pi \omega^2 ut \over s^2} \, , 
\end{equation}
\begin{equation}
p_1\widetilde{M}_{24} p_3 = p_2\widetilde{M}_{13} p_4  = 
-{16\pi \omega^2 st \over u^2} - {8\pi \omega^2 s \over u} \log\left(-{u\over s}\right)\, , 
\end{equation} 
\begin{equation}
p_1\widetilde{M}_{23} p_4 = p_2\widetilde{M}_{13} p_4  = 
-{16\pi \omega^2 su \over t^2} - {8\pi \omega^2 s \over t} \log\left(-{t\over s}\right)\, . 
\end{equation}

Combining with the relevant pre-factors and adding the ${\cal V}$ terms one finds 
\begin{equation}
- {1\over 8}{8\pi G \over 2(2\pi)^3} \int {d^3q \over |q|} {{{\cal U}} + {\cal V}\over \prod_i qp_i} =  - 64 \pi\omega^2 \times  {1\over 8} {8\pi G \over 2(2\pi)^3} \nonumber \end{equation}
\begin {equation}
\left\{ 1 + 2 +\left[ut \left({1\over u^2}-{1\over t^2}\right)
+2st \left({1\over s^2}-{1\over t^2}\right)+2su \left({1\over s^2}+{1\over u^2}\right)  \right. \right.
\end{equation}
$$\left. \left.+ su \left({1\over s^2}-{1\over t^2}\right) \log\left(-{u\over s}\right) + st \left({1\over s^2}+{1\over u^2}\right)\log\left(-{t\over s}\right)\right] \right\}  \, . $$

The final expression can  be rewritten as  \begin{equation}
- {1\over 8} {8\pi G \over 2(2\pi)^3} \int {d^3q \over |q|} {{{\cal U}} + {\cal V}\over \prod_i qp_i} = 
{4G\hbar^2\omega^2\over \pi } \times \end{equation}
$$\left\{- 1 +\left[{s\over t} - {s\over u}  + us  \left({1\over t^2} - {1\over s^2}\right)\log\left(-{u\over s}\right)
- st  \left({1\over s^2}+{1\over u^2}\right)\log\left(-{t\over s}\right)\right] \right\}\, .
$$
This coincides with (\ref{fluxtree}), (\ref{AdaaA}) apart from a change of sign in the constant $+1$ in (\ref{AdaaA}). However, as already mention there, keeping into account the ${{\cal O}}(\omega^2)$ difference between (\ref{sumij}) and (\ref{sumij1}) precisely switches that sign and yields (\ref{AdaaB}). 
Additional additive constants 
could have come from ambiguities in the ordering of the differential operators acting on the amplitudes. The exact agreement seems to indicate that the ordering adopted in this paper is correct.

\end{document}